\documentclass[preprint,,trackchanges]{aastex7}

\usepackage{subfigure}
\usepackage{makecell}
\usepackage{booktabs}
\begin{document}

\title{Investigating the Influence of Radiative Feedback in Bright-Rimmed Cloud~44}

\author[0009-0007-6397-3220]{Rishi C$^*$}
\affiliation{Aryabhatta Research Institute of Observational Sciences (ARIES), Manora Peak, Nainital 263002, India}
\affiliation{Mahatma Jyotiba Phule Rohilkhand University, Bareilly, 243006, U.P., India}
\email[show]{rishictk@gmail.com}  

\author[0000-0002-0151-2361]{Neelam Panwar$^{**}$} 
\affiliation{Aryabhatta Research Institute of Observational Sciences (ARIES), Manora Peak, Nainital 263002, India}
\email[show]{neelam\_1110@yahoo.co.in}
\author[0000-0002-9593-7618,sname='Thomas J. Haworth']{Thomas J. Haworth}
\affiliation{Astronomy Unit, School of Physics and Astronomy, Queen Mary University of London, London E1 4NS, UK}
\email{}
\author[0000-0002-3904-1622,sname=Yan Sun]{Yan Sun}
\affiliation{Purple Mountain Observatory, Chinese Academy of Sciences, Nanjing 210033, People’s Republic of China}
\email{}
\author[0000-0001-5731-3057,sname=Saurabh Sharma]{Saurabh Sharma}
\affiliation{Aryabhatta Research Institute of Observational Sciences (ARIES), Manora Peak, Nainital 263002, India}

\email{}
\author[0000-0002-6740-7425,sname=R. K. Yadav
]{R. K. Yadav
}
\affiliation{National Astronomical Research Institute of Thailand (Public Organization), 260 Moo 4, T. Donkaew, A. Maerim, Chiangmai 50180, Thailand}
\email{}
\author[0000-0001-9312-3816,sname=D.K. Ojha]{D.K. Ojha}
\affiliation{Tata Institute of Fundamental Research, Mumbai 400 005, India}
\email{}
\author[0000-0001-6802-6539,sname=H.P. Singh]{H.P. Singh}
\affiliation{Department of Physics \& Astrophysics, University of Delhi, Delhi—110007, India}
\email{}

\author[0000-0003-4908-4404,sname=Jessy Jose]{Jessy Jose}
\affiliation{Department of Physics, Indian Institute of Science Education and Research Tirupati, Yerpedu, Tirupati - 517619, Andhra Pradesh, India}
\email{}
\author[0000-0001-7881-7748,sname=Ajay Kumar Singh]{Ajay Kumar Singh}
\affiliation{Department of Applied Physics/Physics, Bareilly College, Bareilly, 243006, India}
\email{}
\author[0009-0005-5242-3931,sname=Jincen Jose]{Jincen Jose}
\affiliation{Aryabhatta Research Institute of Observational Sciences (ARIES), Manora Peak, Nainital 263002, India}
\affiliation{Center for Basic Sciences, Pt. Ravishankar Shukla University, Raipur, Chhattisgarh 492010, India}
\email{}

\author[sname=Shubham Yadav]{Shubham Yadav}
\affiliation{Aryabhatta Research Institute of Observational Sciences (ARIES), Manora Peak, Nainital 263002, India}
\affiliation{ Indian Institute of Technology Roorkee, Roorkee, 247667, Uttarakhand, India}

\email{}

%\collaboration{all}{The Terra Mater collaboration}

\begin{abstract}

Radiative feedback from massive stars plays a central role in the evolution of molecular clouds and the interstellar medium. This paper presents a multi-wavelength analysis of the bright-rimmed cloud, BRC 44, which is located at the periphery of the H{\sc ii} region Sh2-145 and is excited by the massive stars in the region. We use a combination of archival and newly obtained infrared data, along with new optical observations, to provide a census of young stellar objects (YSOs) in the region and to estimate stellar parameters such as age, mass etc. The spatial distribution of YSOs visible in the optical wavelength suggests that they are distributed in separate clumps compared to the embedded YSOs and are relatively older. Near-Infrared (NIR) spectroscopy of four YSOs in this region using the TANSPEC mounted on the 3.6m Devasthal Optical Telescope (DOT) confirms their youth. 
From Spectral Energy Distribution (SED) fitting, most of the embedded YSO candidates are in their early stage of evolution, with the majority of them in their Class II and some in Class I stage. The relative proper motions of the YSOs with respect to the ionizing source are indicative of the rocket effect in the BRC. The $^{12}$CO, $^{13}$CO, and C$^{18}$O observations with the Purple Mountain Observatory are used to trace the distribution of molecular gas in the region. A comparison of the cold molecular gas distribution with simple analytical model calculations shows that the cloud is in the compression stage, and massive stars may be influencing the formation of young embedded stars in the BRC region due to radiative feedback.

\end{abstract}



\section{Introduction} 

Radiative feedback refers to the influence of the radiation of massive stars on their surrounding interstellar medium (ISM), which consists of gas and dust \citep{2005pcim.book.....T,2015NewAR..68....1D, hopkins20}. This complex process plays a fundamental role in shaping the stellar mass spectrum, as well as the evolution and characteristics of molecular clouds in associated star-forming regions (SFRs) \citep{2009AJ....138....7M}. The interaction between stellar radiation and the surrounding medium can significantly impact star formation within individual SFRs and galaxies by either enhancing or suppressing the process. 
%It could subsequently alter the shape of the Initial Mass Function (IMF), circumstellar disk evolution followed by planet formation \citep{2020MNRAS.492.5030H}.

The high-energy photons from massive stars ionize and heat the surrounding ISM \citep[e.g.][]{2017MNRAS.466.3293P} which leads to the formation of H{\sc ii} regions, expanding ionization/ shock fronts, and the subsequent re-distribution of gas and dust within the SFR \citep[e.g.][]{2009ApJ...692..382M, 2009ApJ...694L..26G, 2012MNRAS.427..625W, 2014MNRAS.442..694D, 2022ApJ...941..202R}. This feedback has a dual impact on star formation. In negative feedback, intense radiation pressure and photoionization of massive stars can disperse surrounding gas and dust, reducing the available material for further star formation, thereby reducing the overall star formation efficiency \citep{2018ApJ...859...68K}. Conversely, under specific conditions, radiative feedback may also compress the surrounding gas, which then promote the formation of new stars \citep{2005MNRAS.358..291D}. One of the scenarios of positive feedback is radiation driven implosion (RDI) \citep{2009ApJ...692..382M, 2011ApJ...736..142B, 2011MNRAS.412.2079M, 2012MNRAS.420..562H}, where the H{\sc ii} region created by the massive stars triggers star formation in nearby pre-existing clumps as an expanding shock front followed by an ionizing front propagates through it \citep{2004A&A...426..535M,2011ApJ...736..142B,2012PhDT.........4T}. Ultimately, reality is a complicated mixture of positive and negative feedback \citep[e.g.][]{2015MNRAS.450.1199D}, and the dynamic interplay between these competing effects determines the evolutionary trajectory of star-forming regions and contributes to the observed diversity in galactic star-formation rates \citep{2009MNRAS.398..157H}.

One of the most recognizable structures associated with radiative feedback is bright-rimmed clouds (BRCs). These regions often exhibit triggered star formation, where the external radiation pressure induces gravitational collapse, leading to the formation of new generations of stars \citep{1989ApJ...342L..87S,2004A&A...414.1017T, 2004A&A...426..535M,2006A&A...450..625U, 2016AJ....151..126S}. 
%These regions serve as crucial laboratories for studying the role of radiative feedback in initiating star formation.

%A possible consequence of triggered star formation is sequential star formation \citep{2017MNRAS.468.2684P,2010ApJ...713..883B}. 
Despite extensive theoretical and observational studies, the precise role of radiative feedback in regulating star formation remains a subject of debate \citep{2016ApJ...822...49J}. As an H\,\textsc{ii} region expands, it interacts with the surrounding molecular cloud, disperses low-density gas, and then compresses denser regions, leading to an observable age gradient among the newly formed stars. Typically, older stars are found closer to the ionizing source, while younger stars are located farther away \citep{2002ApJ...565L..25S,2007ApJ...654..316G,2012ApJ...755...20S}. However, not all observations support this model. \citet{2007ApJ...666..321I} found cases where young stars are randomly distributed rather than following a sequential age gradient. \cite{2015MNRAS.450.1199D} also demonstrated with simulations that the stars that most likely were suggested in observations to be ``triggered'' still formed in control simulations without feedback. This inconsistency has sparked debate over the effectiveness of radiative triggering mechanisms in star formation.

In the present work, we investigated BRC~44 which is a part of H{\sc ii} region Sh2-145 \citep{1995yCat..41130325H} and is located at a distance of $\sim 910$\,pc \citep{1974PDAO...14..283C,1995yCat..41130325H}. BRC 44 is an active
site of recent/ ongoing star formation. It harbors the dark cloud LDN 1206, which consists of two IRAS sources: IRAS 22272+6358A (hereafter `source A') and IRAS 22272+6358B (hereafter `source B').
%Many $H\alpha$ emission-line stars are found at the edge  of the cloud\citep{2002AJ....123.2597O}. 
Spitzer-IRAC and MIPS observations of the region reveal an embedded cluster of young stellar sources inside the BRC. \citet{2009ApJS..184...18G} identified three Class {\sc i} and twenty-two Class {\sc ii} sources in the region.  
 A wide $H\alpha$ observation by \citet{2002AJ....123.2597O} revealed an elongated distribution of the emission stars near the head of the BRC and a small reflection nebula near the western part of the BRC 44. They also found Herbig Haro HH 596, a delicate jet-like object that runs through the symmetry axis of the reflection nebula. The position of IRAS 22266+6358 coincides with it, so it was identified as the probable exciting source of the HH 596.
Recently, \citet{2020ApJ...904...75L} investigated a sample of 14 intermediate-mass protostars selected from the SOFIA Massive Star Formation Survey (SOMA), including the IRAS sources associated with BRC 44 (source A and source B). They performed spectral energy distribution (SED) fitting to these sources using Spitzer, %$Herschel$, 
SOFIA and IRAS fluxes and extracted the physical parameters like mass, bolometric luminosity etc. %The result obtained by them from SED fitting is compared with our result in discussion section.
%We studied each YSO candidate obtained from NIR and MIR imaging using color-color and color-magnitude diagrams and later fitted SEDs to the sources to constrain their physical parameters.
%Medium Resolution Spectroscopy of selected YSOs will help us say at what stage each YSO is in right now, whether they follow with the results obtained from the photometric study, etc. 
%\citet{2004A&A...426..535M} using NRAO/VLA Sky Survey (NVSS) calculated external pressure by (IBL) on the cloud for different BRCs in which BRC - 44 was one of the candidates. Using the line ratios of the CO isotopologues and the study mentioned above, we will get an overall picture of how the BRC is influenced by external radiation.
%The source IRAS 22198 + 6336A was not detected and classified as YSO in our study since it has no optical counterpart and is detected only as scattered light in NIR wavelengths\citep{1991AJ....102.1398R}. It was identified as an intermediate-mass protostar by \citet{2020ApJ...904...75L}.
CO(1–0) observations of source A by \citet{2006A&A...457..865B} reveal that a collimated outflow is driven by this source
with a weak southeastern red lobe and a much stronger northwestern blue lobe. They concluded that the weak red lobe may be due to the erosion of the outflow by the expanding H{\sc ii} region due to photoionization by a nearby massive star. 

\begin{figure*}

\includegraphics[width = \textwidth]{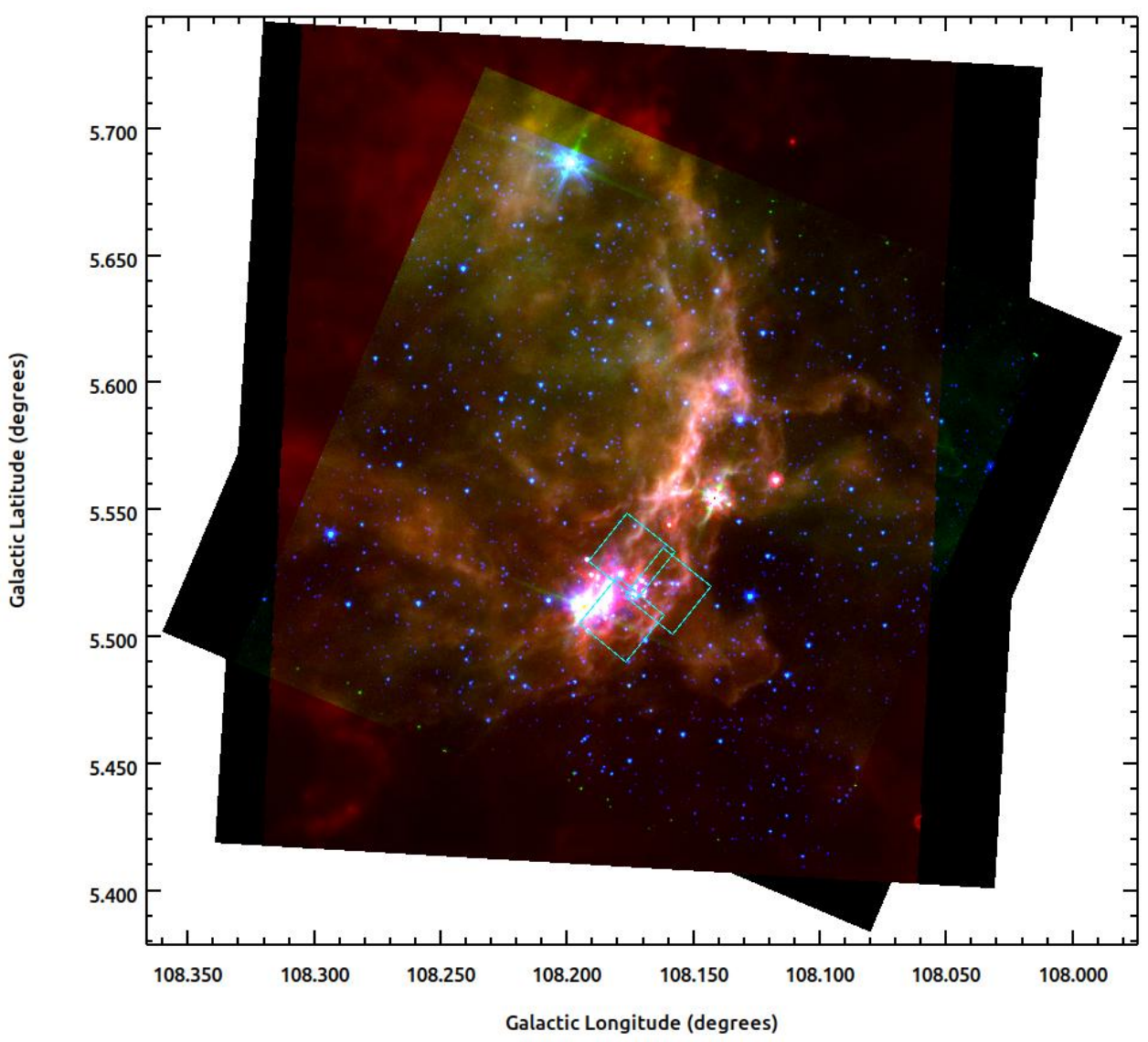}
\caption{RGB image of the BRC~44 using Spitzer IRAC 3.6\,$\mu$m (Blue), Spitzer IRAC 5.8\,$\mu$m (Green), and Spitzer MIPS 24\,$\mu$m(Red) data. The area observed using TIRCAM2 is shown using boxes.}
\label{fig:tircam}
\end{figure*}

Here, we present a multi-wavelength analysis of BRC~44 (see. Figure \ref{fig:tircam}) to identify the young stellar population and the possible role of feedback in their formation through optical to radio imaging and spectroscopic observations. 
The structure of the paper is as follows. Section 2 summarizes the observations and archival data used for the work. Section 3 describes the results and analysis of the present work, including the identification of the young stellar objects (YSOs) and the derivation of their physical parameters, as well as the study of the structure and parameters of the cloud. Section 4 discusses our results related to the triggered star formation scenario. In the last section, we conclude our work by highlighting our results.

\section{Data used}
\subsection{Observations and data reduction}
\subsubsection{Optical $V$ and $I$ data}
We observed BRC~44 in optical $V$ and $I$ bands using a 2K $\times$ 2K ANDOR CCD camera(FOV of 18$'$ $\times$ 18$'$) mounted on the 1.3-m Devasthal Fast Optical Telescope (DFOT) \citep{2011CSci..101.1020S, 2022JAI....1140004J}on November 15, 2023. The FOV for the imager covers the required target FOV. We also observed a Landolt standard field, SA~92, taken from the European Southern Observatory standard star catalog on the same night to perform photometric standardization. The standard field was observed at different airmasses from 1.3 to 2.1. %Photometry was performed for both standard and target fields.
%The extinction correction was done using the standard field in both $V$ and $I$ bands. The values of extinction coefficients for $V$ and $I$ were 0.178 and 0.068, respectively. 
A linear fit was carried out for the standard star to obtain the color and magnitude transformation equations, and these transformation coefficients were then applied to the target field to obtain the standard $V$ and $I$ magnitudes of all the detected sources with magnitude uncertainties less than 0.1 mag in the region \citep{1992JRASC..86...71S}. The photometric depth obtained through this optical observation in $V$ band was $\sim$ 21 mag and that of $I$ band was $\sim$ 18.7 . 
\subsubsection{Near-Infrared Imaging}
The photometric near-infrared (NIR) $JHK$ observations of BRC 44 were taken on October 15, 2020, using a 512×512 pixel imaging camera, TIFR Near-Infrared Imaging Camera - 2 (TIRCAM2), mounted
on the 3.6 m Devasthal Optical Telescope (DOT). The region covered is shown in Figure \ref{fig:tircam}. Observation is taken in three pointings with a Field of View (FOV) of single pointing 86.5$''$ $\times$ 86.5$''$. TIRCAM2 provides NIR imaging observations in the wavelength range of 1 to 3.7$\mu$m. Detailed information on the instrument is given in \citet{2018JAI.....750003B}. The details of the observation are given in Table \ref{table1}. The typical seeing size during the observations was about $0.5''$. As our
TIRCAM2 observations in the $JHK$ bands cover only a small portion of the BRC 44 region; we also used the two-micron all-sky survey (2MASS) data for the wider area. In addition to this, we have also used 2MASS measurements of the bright sources that were saturated in our deep $JHK$ observations. Thus, we have made a combined catalog of sources detected in the $JHK$ bands using the 2MASS and TIRCAM2 data.
Further, we considered only those sources that have magnitude uncertainty $\leq$0.1 mag in the $J$ and $ H$ bands. The photometric depth obtained through NIR observation in $J$ band was $\sim$ 20.4, that of $H$ band was $\sim$ 22.1 and that of $K$ band was $\sim$ 17.
\subsubsection{Near-infrared spectroscopy of selected YSOs}
We carried out NIR spectroscopy of four YSOs (see Section \ref{res}) using TIFR-ARIES Near Infrared Spectrometer (TANSPEC). The details of the observation are given in Table \ref{table1}. TANSPEC is a medium-resolution spectrograph and camera with sensitivity in the wavelength range from 550 to 2540 nm \citep{2022PASP..134h5002S}. The observations were carried out in cross-dispersed (XD) mode with R $\sim$ 2750 and a slit width of 0.5''. The telluric standard star was used from the ESO telluric standard catalog for telluric line corrections. Data reduction is carried out using $PYTANSPEC$ pipeline \citep{2023JApA...44...30G}. After dividing the source spectra with that of telluric standard star, we normalized the obtained spectra using its continuum. 

\subsection{Ancillary Data}
\subsubsection{GAIA DR3 data}
We use Gaia Data Release 3 (DR3) data \citep{2023A&A...674A...1G} to examine the kinematics of the identified YSOs and their distances.  
All the sources within a circular radius of about 18$^\prime$ (similar to FOV of DFOT) centered around LDN 1206 having G-magnitude uncertainty $\le$ 0.1 mag, error in proper motions in RA ($\mu_\alpha$) and Dec ($\mu_\delta$) $\le$ 0.5 and RUWE $\le$ 1.4 were selected for the analysis.
\subsubsection{Spitzer-IRAC and MIPS Observation}
We obtained processed and mosaicked Spitzer Enhanced Imaging Products (SEIP) images
in IRAC and MIPS bands from the Spitzer heritage archive. The Spitzer-IRAC and MIPS images
were taken at wavelengths of 3.6 $\micron$ (Ch1), 4.5 $\micron$ (Ch2), 5.8 $\micron$ (Ch3), 8.0 $\micron$ (Ch4) and 24 $\micron$ (M1). IRAC images were taken in high dynamic range mode, i.e., both short- and long-exposure images were available. 
%We carried out photometry on both the short- and long-exposure images separately. The SEIP images were in units of MJy/sr, so they were converted to analog-to-digital units (ADU) for photometry. The conversion factor was given in the IRAC and MIPS handbooks and in the image headers.
We used the $DAOPHOT$-II \citep{Stetson:1987iu} package available in the IRAF software to perform point-spread function (PSF) photometry on the fits images. 
Photometry of the bright stars, which were saturated in the long-exposure frames, was taken
from the short-exposure frames. The zero-point magnitudes were taken from the IRAC data handbook to obtain standard magnitudes of the detected sources in all Spitzer bands.

%After photometry is performed in all IRAC bands and MIPS 24 $\micron$, we developed a Python program to merge photometry in all wavelength bands. We have adopted a matching radius of $\sim$ $1''$ for band merging our long/ short IRAC and MIPS catalogs. The NIR $J$, $H$, $K$ data of the region observed using 2MASS/ TIRCAM2 were also merged with the Spitzer-IRAC and MIPS data using the same matching radius.

\begin{table*}
    \centering
    \caption{Log of observations}
    \label{table1}
    
    % First part: Imaging
    \begin{tabular}{cccccc}
        \toprule
        \multicolumn{6}{c}{\textbf{NIR Imaging (TIRCAM2)}} \\
        \midrule
        Date of Observation & Filter & Exp Time (s) & No. of Frames & Dither & No. of Subregions \\
        \midrule
        2020-10-15 & $J$ & 50 & 3 & 5 & 3 \\
        2020-10-15 & $H$ & 50 & 3 & 5 & 3 \\
        2020-10-15 & $K$ & 10 & 15 & 5 & 3 \\
        \bottomrule
    \end{tabular}
    \vspace{1cm}

    % Second part: Spectroscopy
    \begin{tabular}{cccccccc}
        \toprule
        \multicolumn{8}{c}{\textbf{NIR Spectroscopy (TANSPEC)}} \\
        
        \midrule
        Source & Date of Observation & Exp Time (s) & No. of Frames & Dither & $\alpha_{2000}$ & $\delta_{2000}$ & K (mag) \\
        \midrule
        S\_0 & 2023-12-19 & 180 & 3 & 2 & 22:28:57.50 & +64:13:37.5 & 8.25 \\
        S\_1 & 2023-12-19 & 180 & 3 & 2 & 22:28:47.05 & +64:13:14.0 & 9.45 \\
        S\_2 & 2023-12-19 & 180 & 3 & 2 & 22:28:21.13 & +64:14:11.7 & 9.70 \\
        S\_3 & 2023-12-19 & 180 & 3 & 2 & 22:28:43.93 & +64:13:25.7 & 11.51 \\
        \bottomrule
    \end{tabular}
\end{table*}

\subsubsection{CO molecular line data}
We obtained molecular line calibrated data using the 13.7-m millimeter-wave Purple Mountain Observatory (PMO) radio telescope located in Delingha. The data was taken in the $^{12}$CO (J=1–0), $^{13}$CO (J=1–0), and C$^{18}$O (J=1–0) lines along the Galactic Plane.
The $rms$ noise level, which is about 0.5 K for $^{12}$CO (J=1–0) and 0.3 K for $^{13}$CO (J=1–0) and C$^{18}$O (J=1–0), is uniform across the whole covered area. For details on CO data quality, see \citet{2021ApJS..256...32S,2019ApJS..240....9S}.
\section{Analysis and Results}
\label{res}
\subsection{Identification of YSOs}
The identification of member stars in an SFR is helpful in investigating the history of star formation and the influence of massive stars on subsequent star formation in the region. In the present work, we have considered young stellar sources associated with the region as members. To identify candidate YSOs in the region, we have used different approaches using $ Spitzer$ -IRAC / MIPS and NIR data \citep{2009ApJS..184...18G,2017MNRAS.468.2684P}. The detailed procedure is provided in the appendix \ref{identify_spitzer_app}.

\subsection{Physical parameters of the YSOs}

\subsubsection{$V$ vs ($V$-$I$) color-magnitude diagram}

The physical parameters of YSOs can be best constrained using multiple tools like color-magnitude diagrams (CMD), SED fitting, etc. As the contribution of NIR excess is less in shorter wavelengths, optical CMDs help constrain the age and mass of the stars. We searched for the optical counterparts of the YSO candidates in the region using a matching radius of $\sim$ 1$''$ and found about nine sources that have photometry in the $V$ and $I$ bands. We utilize $V$ and $I$ magnitudes of the sources to construct optical CMD as the maximum number of sources were detected in these filters \citep[e.g.,][]{2024AJ....167..106S}. However, to obtain the age and mass of the objects using the CMD, we need to estimate the distance and reddening of the member stars. 

We analyzed the proper motion and parallax values of the young stellar sources in the region using Gaia DR3 data to remove the outliers and distance estimate. The detailed procedure is given in the appendix \ref{distance_app}. 

\begin{figure*}
    \centering
    \subfigure[]{
        \includegraphics[width=0.6\textwidth]{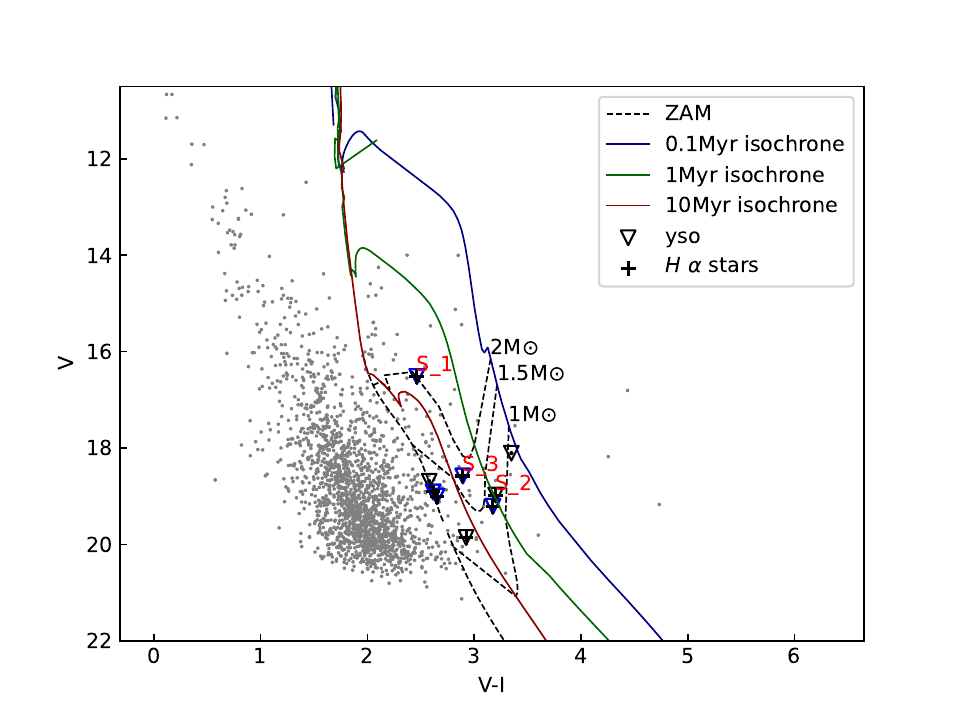}
        \label{fig:vicmd}
    }
    \subfigure[]{
        \includegraphics[width=0.6\textwidth]{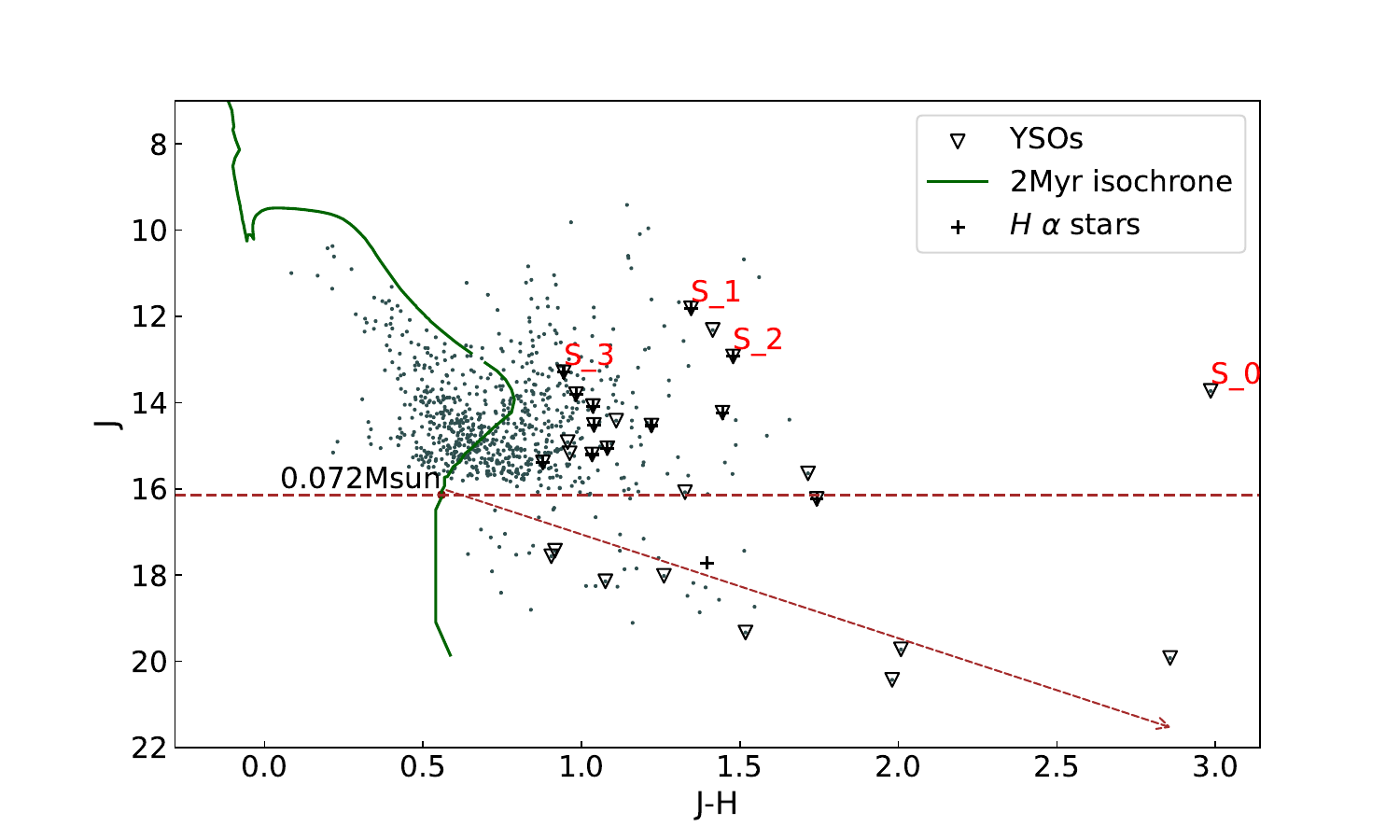}
        \label{fig:jhk_cmd}
    } 
    \caption{The $V$ vs ($V$-$I$) and $J$ vs ($J$-$H$) CMDs for the sources in the BRC 44 region obtained from the DFOT and 2MASS/TIRCAM2 data are plotted in the top and bottom panels, respectively. The inverted triangle markers represent optically visible YSOs identified using NIR and MIR data. Sources $S\_0$, $S\_1$, $S\_2$, and $S\_3$ are marked. Blue inverted triangle markers belong to Group 1 (defined in Section~\ref{identify_spitzer_app}) optically visible YSOs. The brown vector indicates a reddening vector of $A_V = 21$. Note: $S\_0$ is not optically visible.}
    \label{fig:cmds}
\end{figure*}

Using the obtained distance of $\sim 952 \pm 52$, the optical $V$ vs ($V$-$I$) CMD was constructed for the sources in the BRC 44 region, as shown in Figure \ref{fig:cmds} (a). The zero-age main sequence (ZAMS) and pre-main sequence isochrones for 0.1, 1, and 10 Myr age, scaled for the adopted distance and reddening ($A_V \sim 5$ mag), are overplotted on the CMD. %We overplotted 6 million to 7 million years PARSEC isochrones and a zero-age main sequence isochrone to obtain the age and masses of YSOs. The isochrones are distance corrected, 
The reddening was estimated from the average reddening obtained from SED fits of the Group~1 (for details, see Sec. \ref{discuss}) YSOs. All optical YSOs in Group 1 have their $A_V$ between 4 and 7 mag. The evolutionary tracks for the 1, 1.5, and 2 M$\odot$ stars are also shown in the CMD. 
%In addition to the average $A_V$, the individual $A_V$ values of optically visible YSOs derived from the SED fitting were utilized to backtrace the optically visible YSOs to check their age and mass. 
By comparing the location of the YSOs with the theoretical isochrones, we see that most of the Group~1 YSOs are low mass stars with ages of a few Myrs. %They came out similar to the ages and masses obtained from the SED fitting. 
The extinction law used for all the above cases was the interstellar extinction with $R_V= 3.1$ and the relation $E(V-I) = 1.25 \times E(B-V)$ \citep{1989ApJ...345..245C,1979ARA&A..17...73S}. 

\subsubsection{Near-infrared color-magnitude diagram}

As YSOs are deeply embedded in their natal molecular cloud in their early stages, many may not be visible in optical bands. So, NIR photometry was utilized to obtain the physical parameters of the sources not detected in the optical wavelengths. Figure \ref{fig:cmds} (b) shows the $J$ vs. ($J$ - $H$) NIR CMD for the sources in the BRC 44 region. The thick curve overplotted on the CMD represents 2 Myr isochrone from \citet{pastor20} and \citet{1997A&A...327.1054B} for $>$1 M$_\odot$ and $\le$ 1M$_\odot$, respectively, which is scaled for the adopted distance of around $0.95$ kpc. %Some of the masses were also shown along one of the isochrones.

 The identified YSO sources, with available $J$, $H$, and $K$ photometry, are then overplotted to examine their location on the CMD. We observe that these sources are located rightward of the isochrone, indicating that these are reddened sources. We also marked the limit of the brown dwarf mass in Figure \ref{fig:cmds} (b). A reddening vector of A$_V$ $\sim$ 21 mag is also drawn from the brown dwarf (BD) mass limit to show the effect of reddening on the location of the BD candidates on the CMD. The extinction ratios $A_J /A_V$ = 0.265, $A_H/A_V = 0.155$, and $A_K/A_V = 0.090$ were adopted from \citet{1981ApJ...249..481C}. %in the isochrone to remove the interstellar reddening factor from the misidentification source as a brown dwarf. 
 The sources below the reddening vector can be considered BD candidates. Therefore, based on the NIR CMD shown in Figure \ref{fig:cmds} (b), a few sources seem to be BDs. The spatial location of these sources is shown with magenta circles in Figure \ref{fig:brcnew}. %Considering the extinction, we see that most YSOs are less than a million years old. 

 We note that many YSO candidates are grouped around ($J$ - $H$) $\sim$ 1 mag, with most having H$\alpha$ emission. These YSOs seem older than a million years. One of the YSOs, towards the extreme right side in Figure \ref{fig:cmds} (b), seems extremely reddened and relatively massive. This source corresponds to source B (S\_0; in this work) and is discussed in Sec. \ref{discuss} in detail.

 We note that there may be a slight difference in the parameters when considering starspots, specifically using the models provided by Somer et al. (2020). These models account for different surface fractions of starspots on the stars, in contrast to the models that do not include starspots, such as the Baraffe and Parsec models. The models from \citep{2020ApJ...891...29S} estimate the age and mass of stars to be closer to the estimates derived from Baraffe isochrones. However, the CMDs utilized in this study are intended for a rough estimate of the age and mass of the identified YSOs. Also, our primary objective was to compare the age and mass of the two groups of YSOs, and this comparison does not appear to be significantly influenced by the choice of models.

 \subsubsection{Spectral energy distributions of the YSOs}
 \label{sed}
 The SEDs of all YSOs were plotted using SED models from \citet{2006ApJS..167..256R} and the SED fitting tool from \citet{2007ApJS..169..328R} (hereafter R07). These models use the Monte Carlo radiation transfer code to model the radiation transfer. The extinction law of \citet{1994ApJ...422..164K}, which assumes a particle size slightly larger than dust in the interstellar medium, is used in the models. It consists of 14 parameters, including stellar radius, temperature, mass, envelope accretion rate, etc. Using optical $V$ and $I$ photometry, NIR $J$, $H$, and $K$ photometry, and $IRAC$/ $MIPS$ photometry, we performed the SED fitting for all the YSO candidates and extracted their 14 physical parameters.
  We took best-fitting models that satisfied the condition $\chi^{2} - \chi^{2}_{model} < 3\times n_{data}$, where $\chi^{2}$ is the chi-square value of each model, $\chi^{2}_{model}$ is the chi-square value of the best-fitting model and $n_{data}$ is the number of data points available for each source. This criterion is also chosen in R07. We should note that these SED fits give us an estimate of the parameters, and the parameter values obtained may or may not be the actual values. 

\begin{table*}
    \centering
        \caption{Each parameter (extinction $A_V$, age, mass, and accretion rate) is presented with the minimum, best-fit, and maximum values estimated from SED fitting.}
    \label{tab:source_table}
    \begin{tabular}{|c|c|c|c|c|}
        \hline
        \textbf{Source Name} & \textbf{Av} & \textbf{Age(Million years)} & \textbf{Mass($M_{\odot})$} & \textbf{Accretion Rate} \\
        \hline
        S\_0 & \makecell{min: 14.3		 \\ best: 14.3\\max: 22.6} & \makecell{min: 0.2\\best: 5.6\\max: 7.3} & \makecell{min: 4.4\\	best: 6.4\\max: 6.4} & \makecell{min: 1.03e-11		\\	best: 2.20e-10\\max: 2.64e-06} \\
        \hline
        S\_1 & \makecell{min: 4.9	\\ best: 6.0	\\max: 6.0} & \makecell{min: 1.9\\best: 3.2\\max: 3.2} & \makecell{min: 2.7\\	best: 2.7\\max:  3.1} & \makecell{min: 1.19e-08\\	best: 1.19e-08\\max: 1.19e-06} \\
        \hline
        S\_2 & \makecell{min: 4.2\\ best: 4.2\\max: 4.2} & \makecell{min: 0.003\\best: 0.003	\\max: 0.003} & \makecell{min: 0.6\\	best: 0.6\\max:  0.6	} & \makecell{min: 7.29e-06\\	best: 7.29e-06\\max: 7.29e-06} \\
        \hline
        S\_3 & \makecell{min: 6.3 \\ best: 6.6\\max: 6.7} & \makecell{min: 8.2\\best: 8.2	\\max: 9.6} & \makecell{min: 1.8\\	best: 2\\max:  2} & \makecell{min: 5.49e-11\\	best: 5.49e-11\\max: 3.30e-10} \\
        
        \hline
    \end{tabular}

\end{table*}
%\begin{deluxetable*}{ccccc}
% \tablecaption{YSO parameters obtained from SED fitting \label{tab:source_table}}
% \tablewidth{0pt}
% \tablehead{
% \colhead{Source Name} & \colhead{$A_V$} & \colhead{Age (Myr)} & \colhead{Mass ($M_{\odot}$)} & \colhead{Accretion Rate ($M_{\odot}/$yr)}
% }
% \startdata
% S\_0 & min: 14.34, best: 14.34, max: 22.61 & min: 0.2, best: 5.6, max: 7.3 & min: 4.4, best: 6.4, max: 6.4 & min: 1.03e$-$11, best: 2.20e$-$10, max: 2.64e$-$06 \\
% S\_1 & min: 4.87, best: 5.98, max: 6.04 & min: 1.9, best: 3.2, max: 3.2 & min: 2.7, best: 2.7, max: 3.1 & min: 1.19e$-$08, best: 1.19e$-$08, max: 1.19e$-$06 \\
% S\_2 & min: 4.17, best: 4.17, max: 4.17 & min: 0.003, best: 0.003, max: 0.003 & min: 0.6, best: 0.6, max: 0.6 & min: 7.29e$-$06, best: 7.29e$-$06, max: 7.29e$-$06 \\
% S\_3 & min: 6.28, best: 6.64, max: 6.71 & min: 8.2, best: 8.2, max: 9.6 & min: 1.8, best: 2.0, max: 2.0 & min: 5.49e$-$11, best: 5.49e$-$11, max: 3.30e$-$10 \\
% \enddata
% \tablecomments{Each parameter (extinction $A_V$, age, mass, and accretion rate) is presented with the minimum, best-fit, and maximum values estimated from SED fitting.}
% \end{deluxetable*}

\begin{figure*}
    \centering
    \subfigure[]{
        \includegraphics[width=0.45\textwidth]{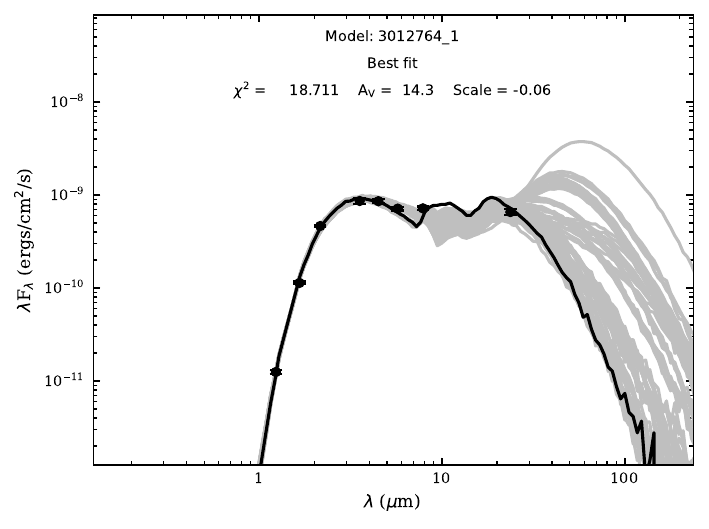}
        \label{fig:sed1}
    }
    \hspace{0.02\textwidth}
    \subfigure[]{
        \includegraphics[width=0.45\textwidth]{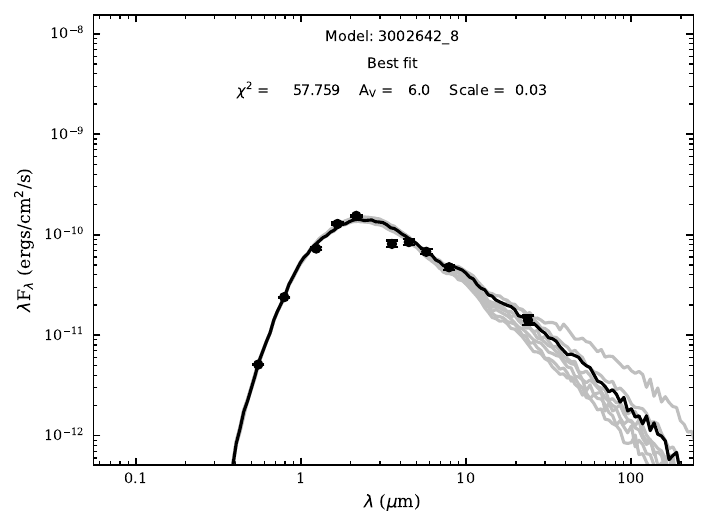}
        \label{fig:sed2}
    }
    \hspace{0.02\textwidth}
    \subfigure[]{
        \includegraphics[width=0.45\textwidth]{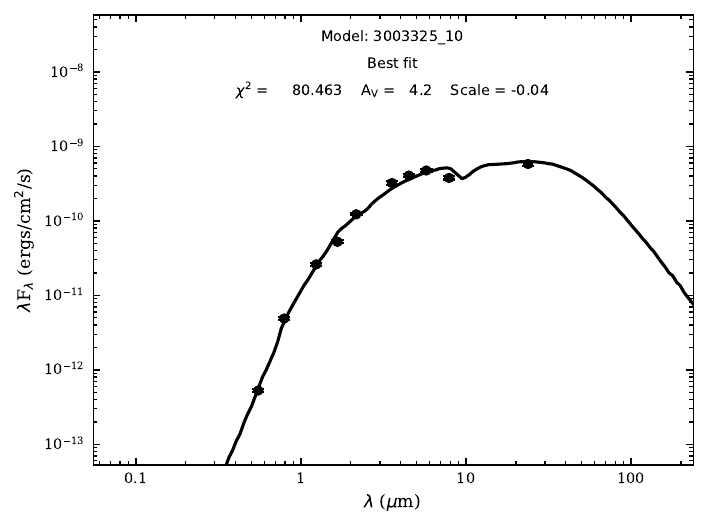}
        \label{fig:sed3}
    }
    \hspace{0.02\textwidth}
    \subfigure[]{
        \includegraphics[width=0.45\textwidth]{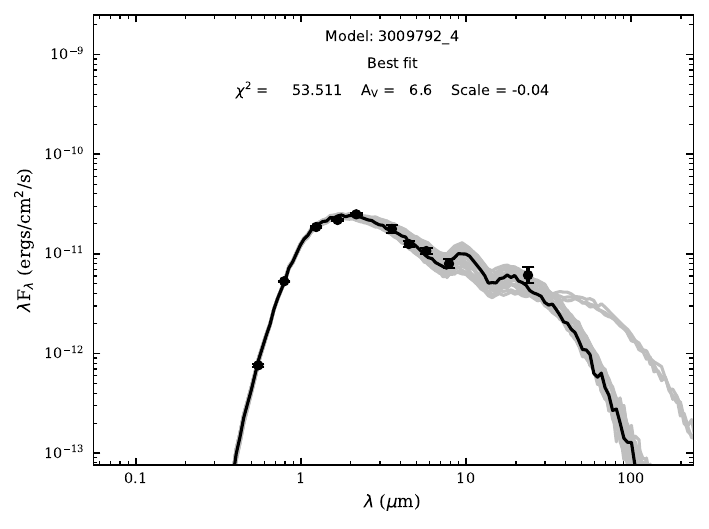}
        \label{fig:sed4}
    }
    \caption{Sample SEDs for the NIR spectroscopic targets. The bold line indicates the best fit with the lowest $\chi^{2}$ value. The light lines indicates fits that has $\chi^{2}$ value following $\chi^{2} - \chi^{2}_{model} < 3\times n_{data}$ criteria.}
    \label{fig:sed_all}
\end{figure*}

 SEDs for the spectroscopic targets are shown in Figure \ref{fig:sed_all} and their physical parameters obtained are given in Table \ref{tab:source_table}. The optically visible YSO candidates (Group 1 sources) have an A$_V$ value comparatively lower than the embedded YSOs that do not have optical counterparts. Almost all YSOs with optical counterparts show the best age with small error bars. Their ages range from 1 Myr to 10 Myr with a mean age of $\sim$ 5 Myr. As seen in Figure \ref{fig:brcnew}, the spatial distribution of these optically visible YSOs (labeled as Group 1) shows that these sources are clumped together and may be part of separate clumps compared to infrared sources.

% \begin{figure*}
%     \includegraphics[width=0.5\textwidth, trim=2.8cm 0 0 0]{star22_spec_ind9.pdf}
%     \includegraphics[width=0.5\textwidth, trim=0cm 0 1.cm 0]{star15_spec_ind6.pdf}
%     \includegraphics[width=0.5\textwidth]{star2_spec_ind9.pdf}
%     \includegraphics[width=0.5\textwidth]{star6_spec_ind7.pdf}
    
% \end{figure*}

 \begin{figure*}
    \centering
    \subfigure[S\_0]{
    \includegraphics[width = 0.45\textwidth]{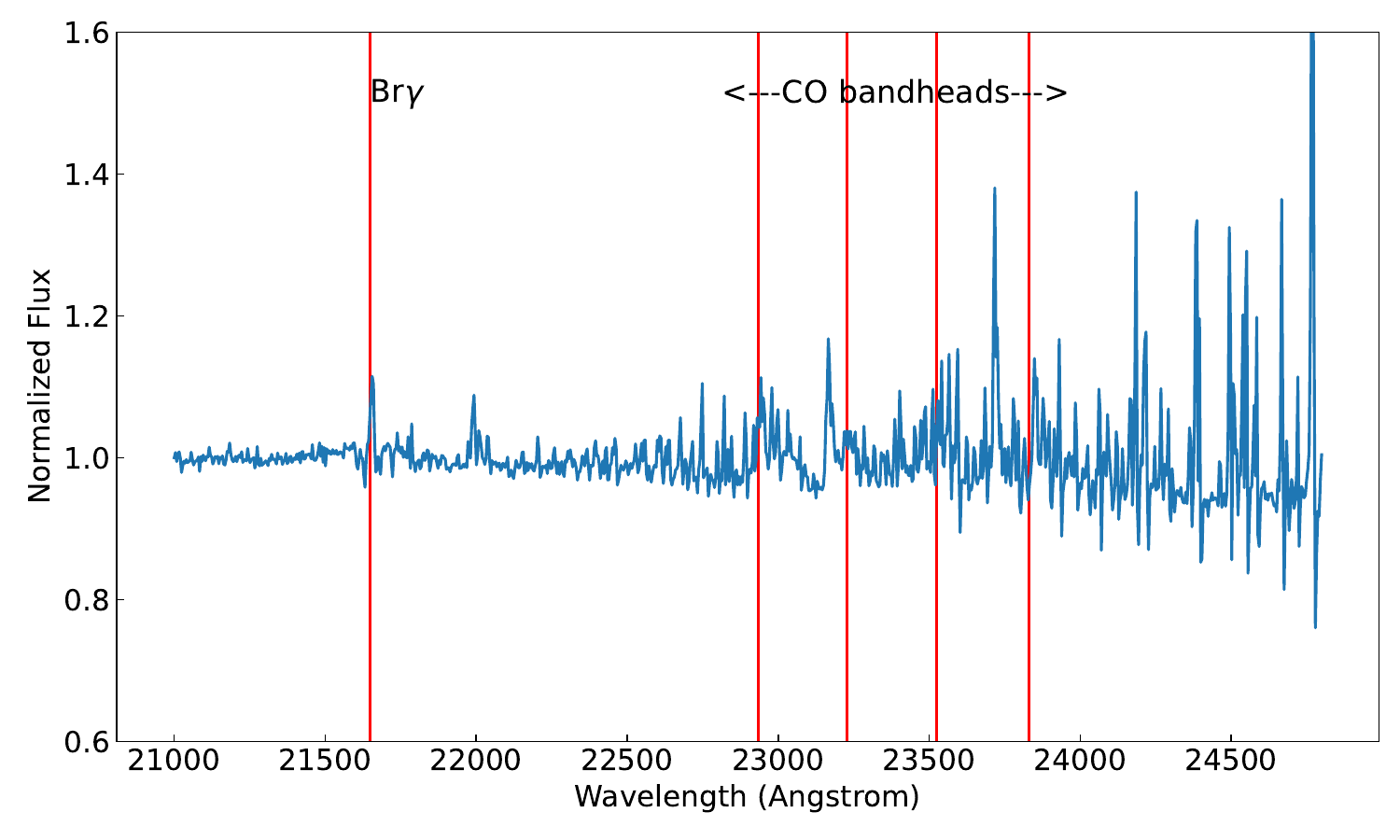}
        \label{fig:spec1}
    }
    \subfigure[S\_1]{
    \includegraphics[width = 0.45\textwidth]{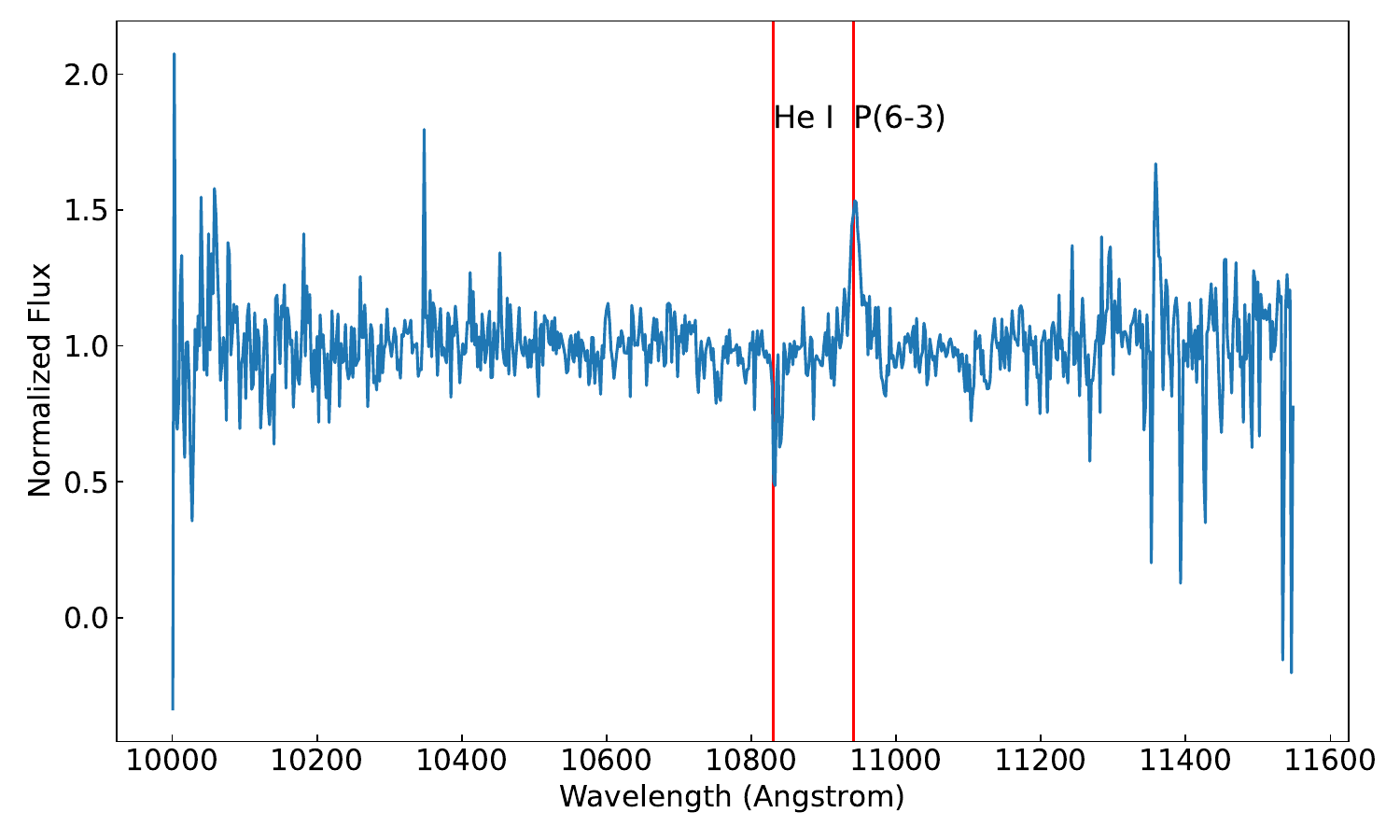}
    \label{fig:spec2}
    }
    \subfigure[S\_2]{
    \includegraphics[width = 0.45\textwidth]{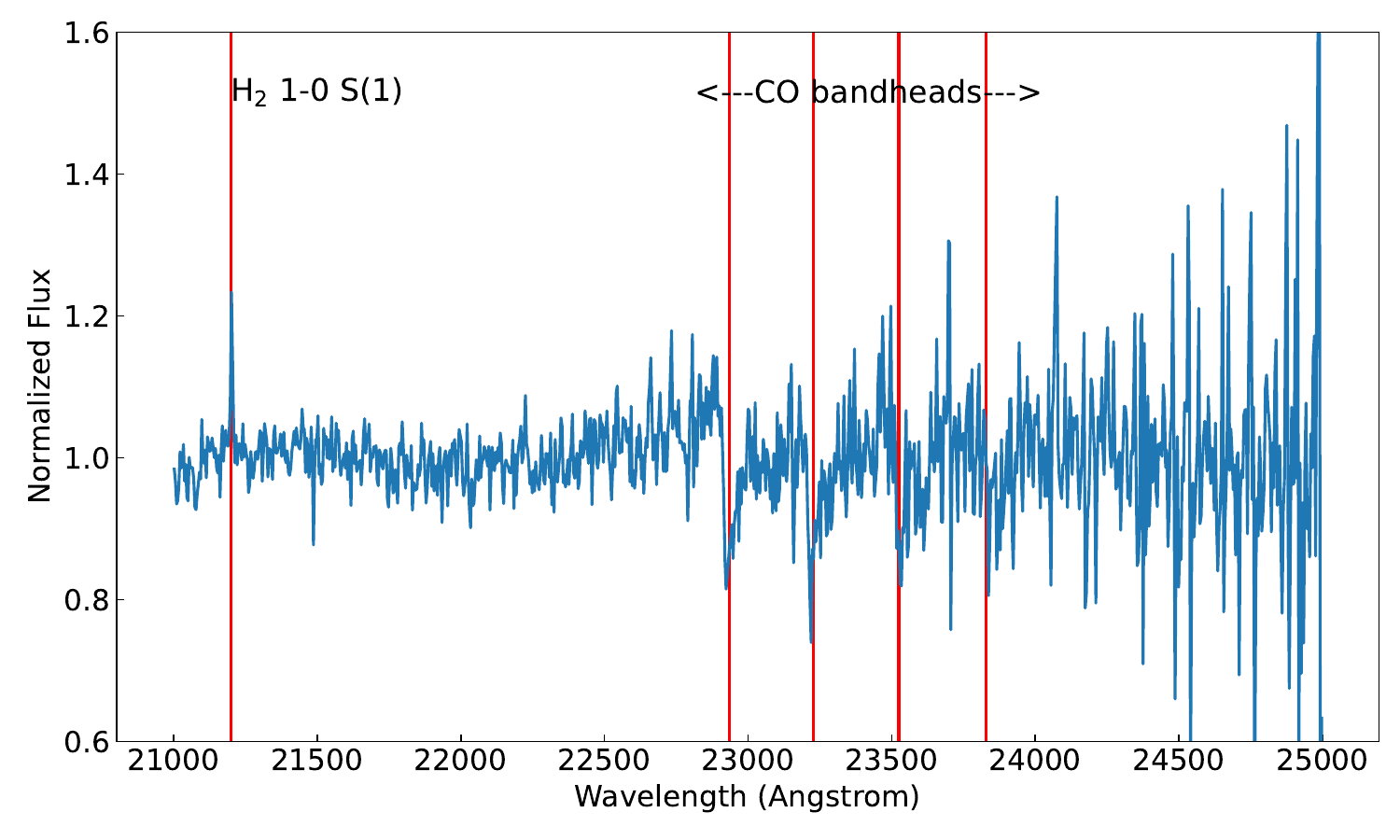}
    \label{fig:spec3}
    }
    \subfigure[S\_3]{
    \includegraphics[width = 0.45\textwidth]{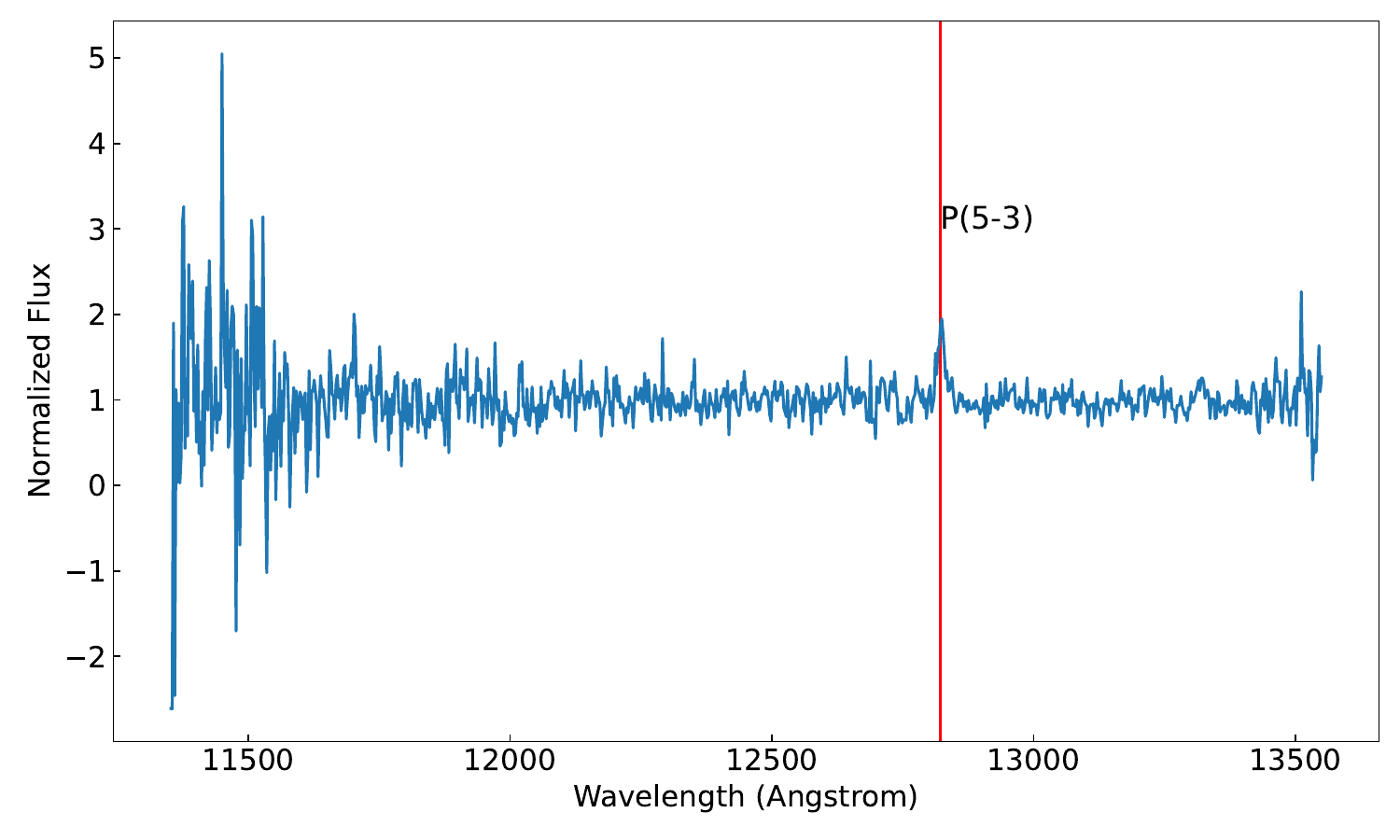}
    \label{image4}
    }
    
    \caption{Wavelength-calibrated normalized NIR spectra of the four sources. \textit{Note:} Lines obtained for each source are in different wavelength ranges. Tanspec covers a larger wavelength range than as shown in figure (5500 to 25400 \AA). We have only plotted regions where features are visible.}
    \label{fig:spectra}
\end{figure*}

\subsubsection{Spectral Analysis of IR bright YSOs}
The spectroscopy of four IR-bright YSOs was performed to study the nature of the sources. The presence of certain absorption or emission lines helps us constrain the mechanism, such as the presence of disk, envelope, accretion, etc. \citep{2010A&A...510A..32M}.
For instance presence of Brackett Gamma (Br$\gamma$) emission line probes the accretion mechanism like magnetospheric accretion, H$_2$ lines probe shocks produced by the outflows from the YSOs, CO $K$-band absorption or emission lines like 2.2935 $\mu \text{m}$ CO(2–1), 2.3227 $\mu \text{m}$ CO(3–1), 2.3535 $\mu \text{m}$ CO(4–2), 2.3829 $\mu \text{m}$ CO(5–3) give the information about the structure of the circumstellar disk, hydrogen recombination lines like Paschen-Beta (Pa$\beta$) and Pa$\gamma$ also probes magnetospheric accretion process, He{\sc i} (1.083 $\mu$m) line indicates the outflowing winds originating from accretion \citep{2010A&A...510A..32M,Ghosh2023}. 
The spectra of the four sources are shown in Figure \ref{fig:spectra}. In the Sec. \ref{discuss}, we discuss the outcomes of the SED fitting with the spectroscopic outcomes of these YSOs.

\subsection{CO data analysis}

\subsubsection{Cloud parameter estimation}
\label{cloud para}
\begin{figure*}

	\includegraphics[width = \textwidth]{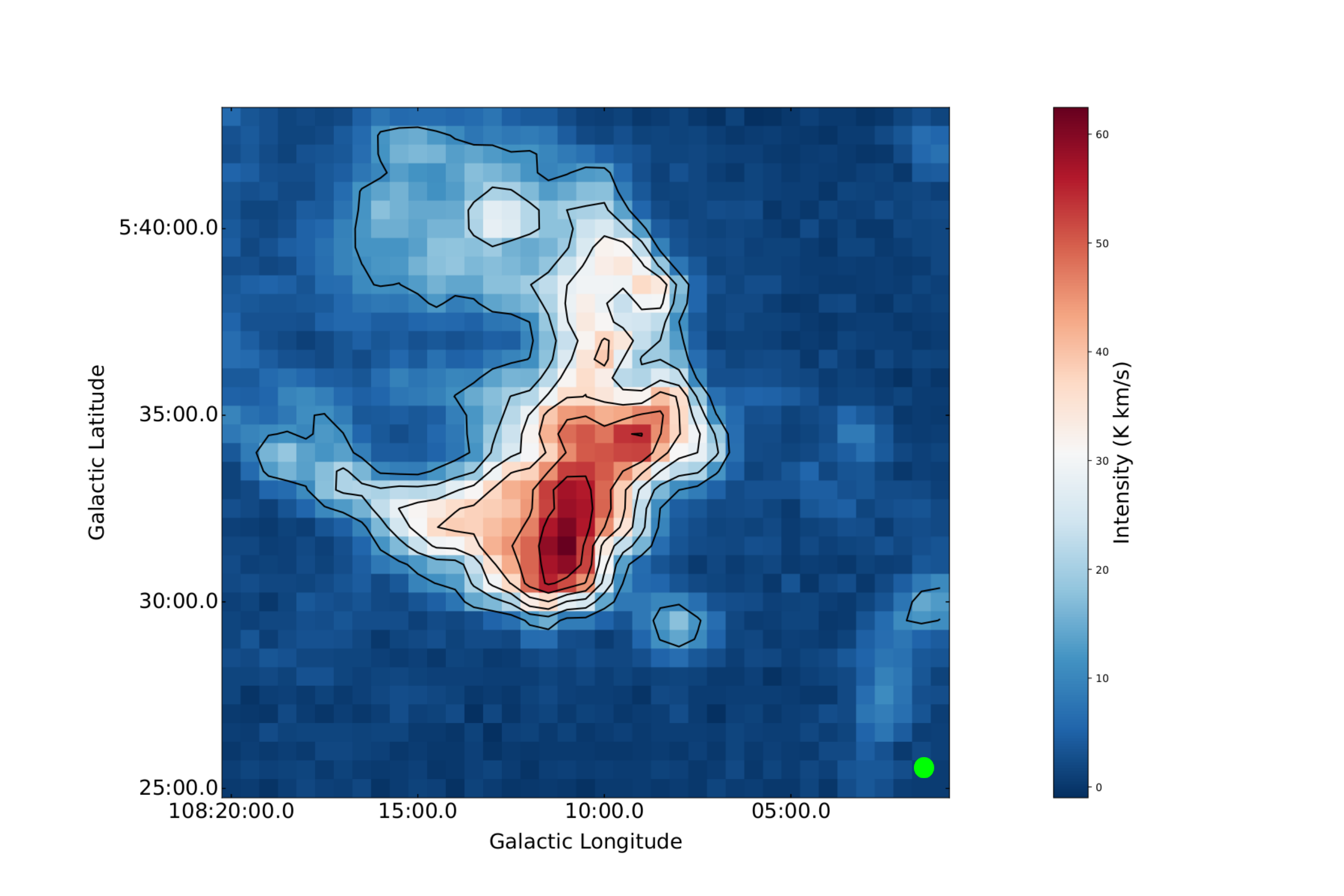}
    \caption{$^{12}$CO contour map of the region. Velocity range is between -13 km s$^{-1}$ and -5 km s$^{-1}$. The contour levels are 11.1, 19.7, 28.2, 36.8, 45.3,
       53.9, 62.5 K km s$^{-1}$. Beam size is shown in bottom right (50$^{\prime\prime}$).}
    \label{fig:12co_moment0}
\end{figure*}

%The above analyses were limited to identifying the YSOs in the region and obtaining their properties. One of the indicators of future SF is the presence of molecular gas in an SFR. Therefore, 
To study the future/ ongoing star formation in the region and the dynamics and structure of the cloud, we probed the molecular material using the millimeter observations obtained from the PMO. %We analyze $^{12}CO$, $^{13}CO$, and $C^{18}O$. Apart from this, 
For the analysis of CO data, we used the Spectral Cube package of Python \citep{2019zndo...2573901G}.
The presence of an optically thick molecular gas is investigated through the observations of $^{12}$CO and optically thin molecular gas is traced using $^{13}$CO and C$^{18}$O \citep{2013A&A...556A.105O}.
%Using $^{12}$CO and $^{13}$CO, we created integrated intensity maps in units of Kelvin km/s as shown in Figure\ref{fig:radiocontour}. 
We identified the cores using optically thin C$^{18}$O emission. The core coincides with the position of the embedded protostar (Figure \ref{fig:brcnew}). We used optically thick $^{12}$CO to analyze the overall structure of the cloud. The contour of $^{12}$ CO traces the cloud structure, as shown in Figure \ref{fig:12co_moment0}.

\begin{figure}
    \centering
    \subfigure[$^{12}$CO spectra]{
        \includegraphics[width = 0.45\textwidth]{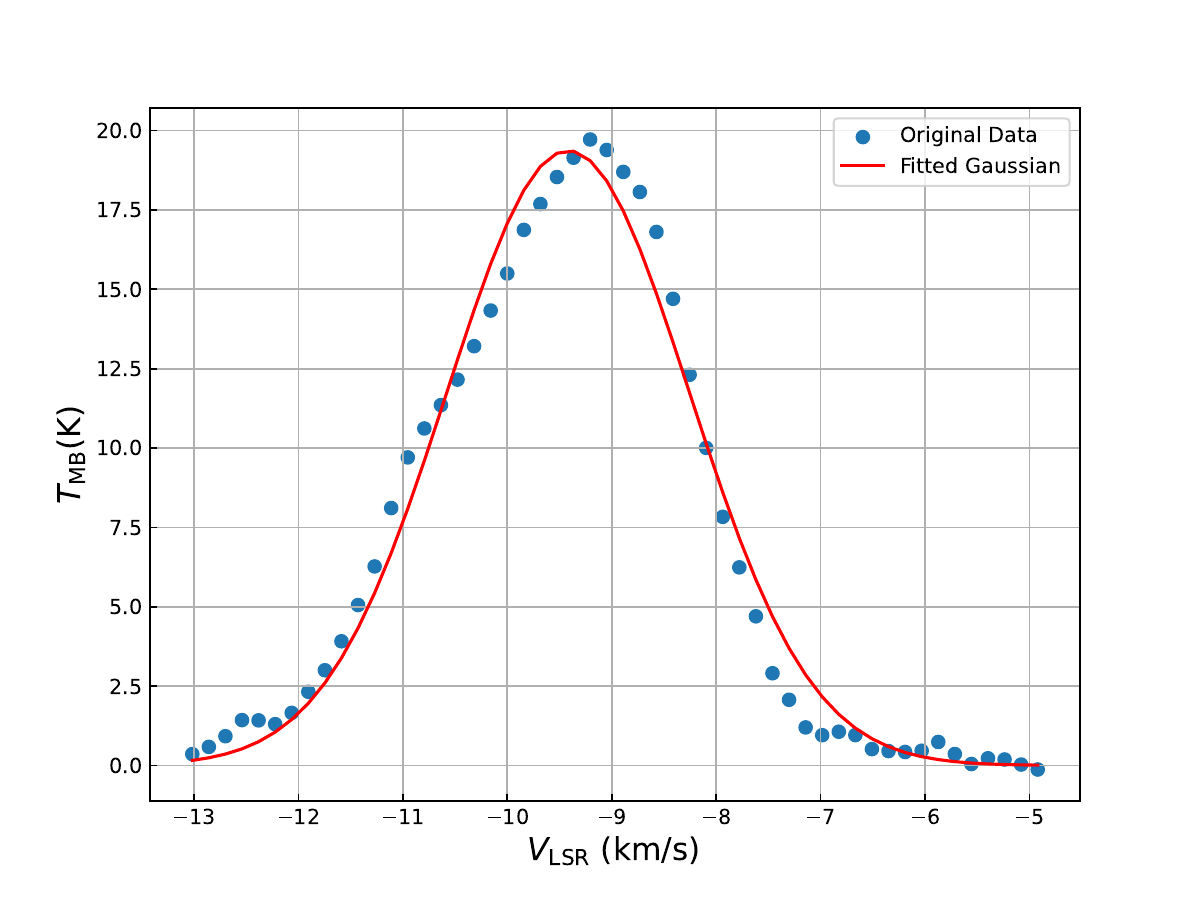}
        \label{fig:radspec1}
    }
    % \hspace{0.05\textwidth}
    \subfigure[$^{13}$CO spectra]{
        \includegraphics[width = 0.45\textwidth]{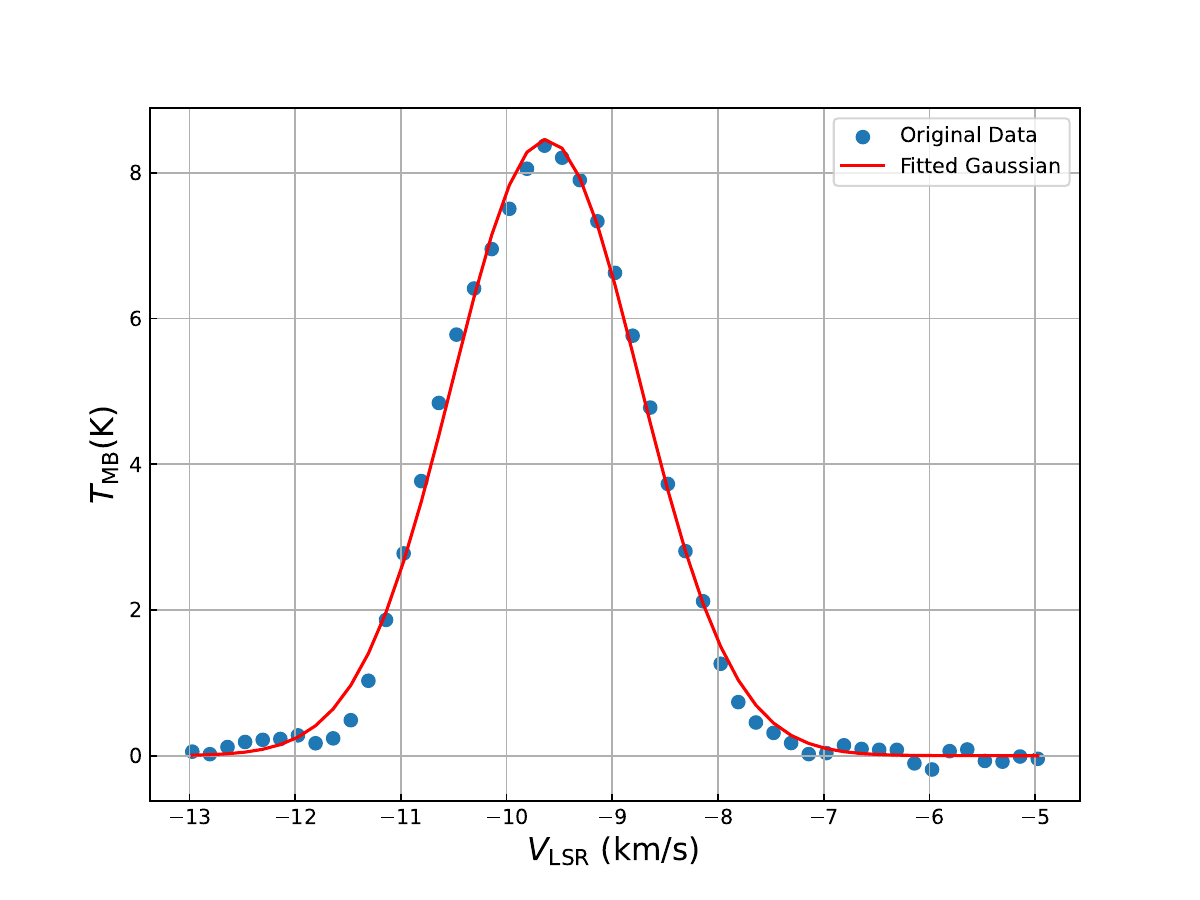}
        \label{fig:radspec2}
    }
    % \hspace{0.05\textwidth}
    \subfigure[C$^{18}$O spectra]{
        \includegraphics[width = 0.45\textwidth]{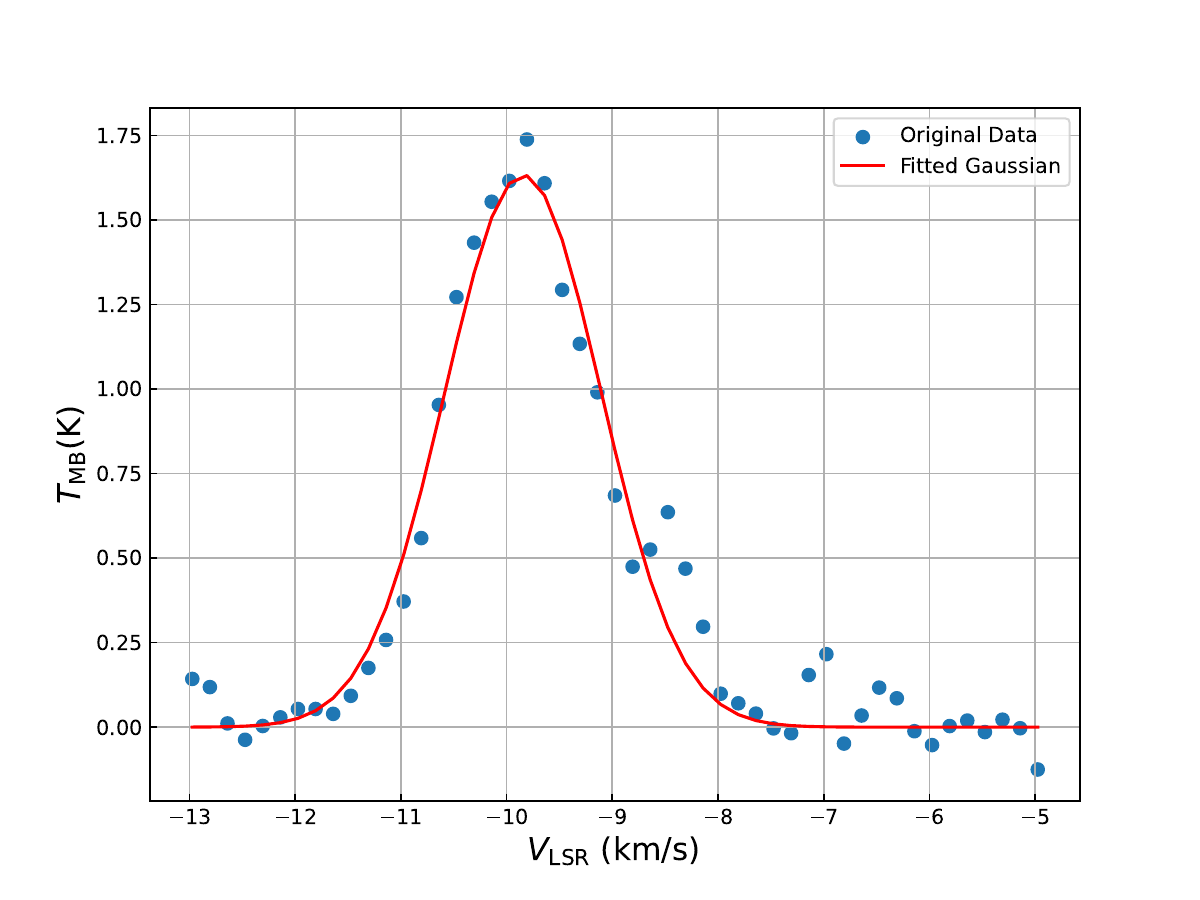}
        \label{fig:radspec3}
    }
    % \hspace{0.05\textwidth}
    
    \caption{Average molecular line spectra for the region. The region used for extracting the spectra is a circular aperture of 42'' as discussed in Section \ref{cloud para}}
    \label{fig:radiospectrum}
\end{figure}

The mean molecular spectra were calculated by Gaussian fitting of the three molecules in the velocity range -15 km s$^{-1}$ to 0 km s$^{-1}$, as shown in Figure \ref{fig:radiospectrum}. All significant features of the cloud are within this velocity range. To examine the impact of the RDI on the star formation activity in BRC, we focused on a circular region with a radius of 42$^{\prime\prime}$ (see, Figure \ref{fig:brcnew}), which corresponds to a physical size of $\sim$ 0.18 parsecs. This region is centered around the protostar source A, near Group 2 in Figure \ref{fig:brcnew}, and believed to be undergoing active star formation as a result of the RDI \citep[see,][]{1989ApJ...342L..87S}. 
A local thermodynamic equilibrium (LTE) radiative transfer equation was modified to derive temperatures and optical depth from \citet{2008ApJ...679..481P}, using the equations given below,
\begin{equation}
I_{\text{line}} = (S  - I_0)(1 - e^{-\tau}),
\end{equation}
where, $I_{\text{line}}$ is the intensity of the emission line. $S$ is the source function, $I_0$ is the intensity of the impinging radiation and $\tau$ is the optical depth. 

The brightness temperature is given by,
\begin{equation}
T_R = \frac{I_\nu c^2}{2 \nu^2 k},
\end{equation}
$I_\nu$ is the specific intensity, $\nu$ is the frequency, and $k$ is the Boltzmann constant.
\begin{equation}
T_R = T_0 \left( \frac{1}{e^{T_0/T_{\text{ex}}} - 1} - \frac{1}{e^{T_0/T_{\text{bg}}} - 1} \right) \left( 1 - e^{-\tau} \right),
\end{equation}
where $\mathrm{T_0} = \frac{\mathrm{h\nu}} {\mathrm{k}}$, ($\frac{h \nu(^{12}\mathrm{CO})}{k} = 5.5\mathrm{K}$ and $\nu(^{12}$CO) = 115.3GHz ), $\mathrm{T_{ex}}$ is the excitation temperature, $\mathrm{T_{bg}}$ is the temperature of the cosmic microwave background which is $2.7$K.

The excitation temperature is given by,
\begin{equation}
T_{\text{ex}} = \frac{5.5 \, \text{K}}{\ln \left( 1 + \frac{5.5 \, \text{K}}{T_{\text{max}}(\mathrm{^{12}CO}) + 0.82 \, \text{K}} \right)} ,
\end{equation}
where, T$_{\text{max}}$($^{12}$CO) is the main beam brightness temperature at the peak of $^{12}$CO.
Assuming that the $^{12}$CO(1-0) excitation temperature is the same as the $^{13}$CO(1-0) excitation temperature, the optical depth of $^{13}$CO(1-0) line can be calculated using the equation
\begin{equation}
\tau(^{13}\mathrm{CO}) = -\ln[1-\frac{T_{max}(^{13}\mathrm{CO})/5.3K}{\frac{1}{(\exp{\frac{5.3K}{T{ex}}-1)}}{-0.16}}] ,
\end{equation}
where, T$_{\text{max}}$($^{13}$CO) is the main beam brightness temperature at the peak of $^{13}$CO. The mean optical depth of $^{13}$CO was found to be $\sim$ 0.5, which is less than 1, so we can consider it marginally optically thin \citep{2021A&A...654A.144B}.

The column density of $^{13}$CO, 
N($^{13}$CO), is given by,

\begin{equation}
\begin{array}{l}
N(\mathrm{^{13}CO}) = \frac{\tau(\mathrm{^{13}CO})}{1 - e^{-\tau(\mathrm{^{13}CO})}} \times 3.0 \times 10^{14} \times W(\mathrm{^{13}CO}) \\
\qquad \times \frac{1}{1 - e^{-5.3 / T_{\text{ex}}}} \, \mathrm{cm}^{-2} ,
\end{array}
\end{equation}

where $W(^{13}$CO) is the $^{13}$CO integrated intensity
and finally using the equations mentioned in \citet{1982ApJ...262..590F}, we estimated the molecular hydrogen column density,
\begin{equation}
\frac{N(\mathrm{^{13}CO})}{N(\mathrm{H_2})} = 1.4 \times 10^{-6} .
\end{equation}

% \clearpage

We estimated the mass of the globule that surrounds the embedded (labeled as Group 2) YSOs. The mass was estimated using,

%$M = {\frac{4}{3} \pi \, (\text{r})^3 \, n_{\text{H}_2} \, \mu_{\text{H}_2} \, m_{\text{H}}}$
\begin{equation}
M = \mu_{\mathrm{H_2}}m_H\int N(\mathrm{H_2})dA ,
\end{equation}
where $\mu_{H_{2}} = 2.8$ is the mean molecular weight of the hydrogen molecule, considering the presence of helium, $m_H$ is the
mass of the hydrogen atom, and A is the area of the cloud.
%and $n_{H_2}$ is the number density of molecular hydrogen obtained by dividing the column density of hydrogen by the radius of the cloud.

The estimated mass of the globule turns out to be $M \sim 81\,\mathrm{M_\odot}$, compared to that found by \citet[][,$\sim 95\,\mathrm{M_\odot}$]{1989ApJ...342L..87S}.

Comparing the internal cloud pressure with the external pressure is used to determine whether a BRC is being compressed \citep[e.g.,][]{2004A&A...426..535M, 2012MNRAS.426..203H}. Here, using the velocity dispersion of the $^{13}$CO data, we calculated the internal cloud pressure,
\begin{equation}
P = \mathrm{n_{H_2}}\sigma^2 ,
\end{equation}
where, $\sigma^2$ is the turbulent velocity dispersion. The dominant component of velocity dispersion is turbulent velocity, since thermal velocity will be extremely small considering the very low temperature of the molecular cloud. 
The pressure estimated is $6.6 \times 10^6 \mathrm{K/cm^3}$, calculated using the $^{13}$CO(1-0) molecular line (in units of $\mathrm{P/k_B}$, P is pressure and $\mathrm{k_B}$ is Boltzmann's constant). The dispersion of turbulent velocity was calculated using both C$^{18}$O and $^{13}$CO(1-0) molecular lines. Both show slightly different values, as shown in Table \ref{tab:radiopara}.  The internal pressure calculated using C$^{18}$O turns out to be $4.7 \times 10^6 \mathrm{K/cm^3}$ for the same circular aperture taken. The value of external pressure calculated using the NVSS data by \citet{2004A&A...426..535M} is $8.5 \times 10^6 \mathrm{K/cm^3}$. Although the internal pressure calculated using both species is different as both of them trace different cloud boundaries, both calculated pressures turned out to be lower than the external pressure. This suggests that the BRC is under-pressured.
%One of them has a pressure value above the external pressure, and the other has a value below the external pressure, both of them do not show a large pressure difference compared to the external pressure. So we cannot conclusively say that a shock is being propagated or the stars being formed here are due to external influence just by pressure analysis. More on this topic will be discussed in the discussion section. 
%So, we see that internal pressure is greater than external, which means the cloud is still over pressured compared to external IBL(Ionisation Boundary Layer), so shock has not yet propagated to the cloud, which points that the stars may not be formed from RDI. 

C$^{18}$O traces the densest part of the cloud and hence gives better estimates of cloud properties. If the source is embedded (as in our case), we can take the velocity dispersion obtained from C$^{18}$O, which is 0.74 kms$^{-1}$ to calculate the pressure. This gives a pressure of $4.7 \times 10^6 \mathrm{K/cm^3}$. \citet{2006A&A...457..865B} had estimated the pressure for the same region to be $3.7 \times 10^6 \mathrm{K/cm^3}$  using $^{13}$CO linewidth. Now, if we consider an error in calculating excitation temperature \citep{2008ApJ...679..481P} and propagate the error, we estimate a relative error of 5.4 $\%$ in calculating the number density of molecular hydrogen. Apart from this, there are errors in estimating velocity dispersion (given the high S/N ratio in the dense part, the Gaussian fitting errors for velocity dispersion are negligible) and radius of the cloud, which are difficult to quantify.
For IBL pressure estimation using NVSS, there are different sources of errors, such as fixing an electron temperature in ionized gas, ionizing flux impinging on the cloud etc.

Therefore, the pressure analysis we have undertaken is indicative of a pressure imbalance in the cloud, and we note that various uncertainties may affect our calculations, such as the selection of the radius of the cloud, calculation of the molecular hydrogen density, intrinsic error in the calculation of excitation temperature, optical depth, etc.

\subsubsection{Structure and Dynamics of the Cloud}
\label{struct and dynamics}
\begin{figure*}
	\includegraphics[width = \textwidth]{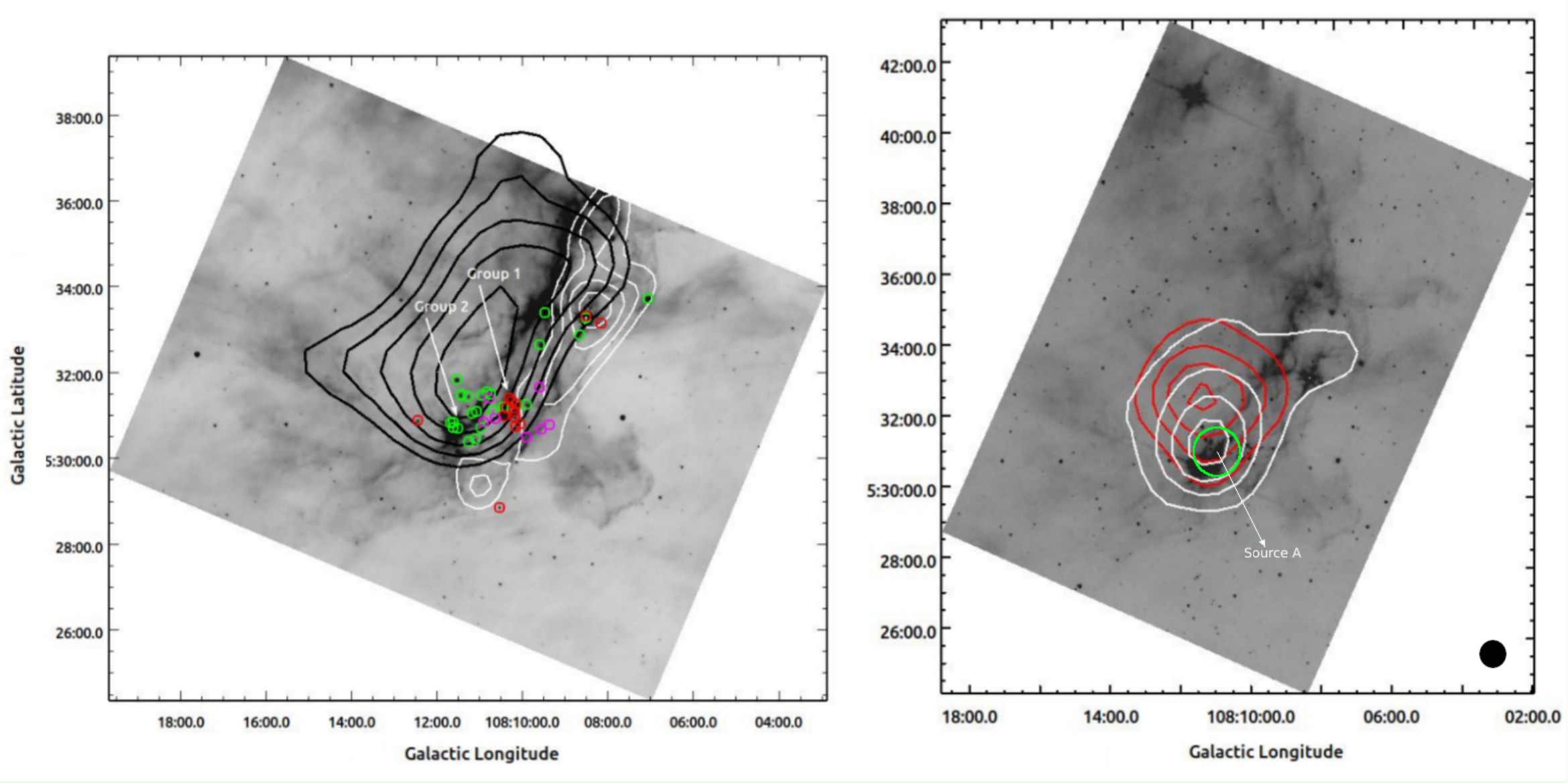}
    \caption{Left: The $^{12}$CO (black color) and 1.4 GHz NVSS (white color) contours are overplotted on the
8 $\mu$m image of the region. Circles represent the identified YSO candidates. The red circles are optically visible YSOs (Group 1), green circles are embedded YSOs(Group 2), and magenta circles are identified as BD candidates. The contour levels for $^{12}$CO are 25, 31, 37, 43, 49 K km $\text{s}^{-1}$, and that of NVSS emission are 0.002, 0.0025, 0.0036, 0.0042 Jy/ Beam.
Right: Two clumps detected from C$^{18}$O (integrated intensity) data overplotted on 5.8$\mu$m image. The white contour contains all the YSOs detected in our study using optical, NIR, and MIR wavelengths. The red contour is the new clump detected, in which no YSOs were detected in our study (velocity range : -9 to -7 $\mathrm{km s^{-1}}$). The contour levels for the red contour are 0.22, 0.37, 0.51, 0.65, and 0.8 K km $\text{s}^{-1}$ and those of the white contour are 0.51 K km $\text{s}^{-1}$, 1.03, 1.55, 2.08, 2.6 K km $\text{s}^{-1}$ respectively. The source A and the circular aperture of 42$^{\prime\prime}$ taken for analysis (green circle) is also shown. Beam size is shown in bottom right (50$^{\prime\prime}$).}
    \label{fig:brcnew}
\end{figure*}

The integrated intensity map using $^{12}$CO emission was generated to visualize the external structure of the molecular cloud, as shown in Figure \ref{fig:12co_moment0}. In Figure \ref{fig:brcnew}, we overlay the NVSS 21 cm radio continuum emission and the $^{12}$CO contour map on the \textit{Spitzer} 8~$\mu$m image. This clearly delineates the molecular cloud’s outer edge and the ionized boundary layer (IBL), traced by the 21 cm NVSS emission. The IBL represents the ionized region along the boundary of molecular cloud that undergoes rapid ionization/ recombination formed by ultraviolet photons emitted from massive stars. Additionally, we present the C$^{18}$O intensity contours overlaid on the \textit{Spitzer} 5.8~$\mu$m image in Figure \ref{fig:brcnew}. The peak of the ${\mathrm{C}^{18}}$O emission (white contour) coincides with the location of a highly embedded protostar belonging to Group 2 YSOs. This spatial correlation indicates active star formation in that region and suggests a relatively young evolutionary stage \citep{2020ApJ...904...75L,1989ApJ...342L..87S}.

\begin{table}
	\centering
	\caption{Parameters obtained from Gaussian fitting of different molecular line spectra for the embedded region.}
	\begin{tabular}{lccr} % four columns, alignment for each
		\hline
		Emission lines & $T_{MB}(K)$ & $V_{LSR}(km/s)$ & $\Delta V(km/s)$\\
		\hline
		$^{12}$CO & 19.37 & -9.42 & 1.16\\
		$^{13}$CO & 8.46 & -9.61 & 0.88\\
		C$^{18}$O & 1.63 & -9.84 & 0.74\\
		\hline
	\end{tabular}
        \label{tab:radiopara}
\end{table}
% \begin{deluxetable}{rlll}
% \tablecaption{Parameters obtained from Gaussian fitting of different molecular line spectra for the embedded region \label{tab:radiopara}}
% \tablewidth{0pt}
% \tablehead{
% \colhead{Emission Line} & \colhead{$T_\mathrm{MB}$} & \colhead{$V_\mathrm{LSR}$} & \colhead{$\Delta V$} \\
% \colhead{} & \colhead{(K)} & \colhead{(km\,s$^{-1}$)} & \colhead{(km\,s$^{-1}$)}
% }
% \startdata
% $^{12}$CO   & 19.37 & $-9.42$ & 1.16 \\
% $^{13}$CO   & 8.46  & $-9.61$ & 0.88 \\
% C$^{18}$O   & 1.63  & $-9.84$ & 0.74 \\
% \enddata
% \tablecomments{Gaussian peak temperatures ($T_\mathrm{MB}$), central velocities ($V_\mathrm{LSR}$), and FWHM line widths ($\Delta V$) are listed for each molecular transition.}
% \end{deluxetable}

%Intensity maps and PV diagrams of $^{12}$CO, $^{13}$CO and C$^{18}$O are created using velocity intervals of $-16$ to $0$ $\mathrm{km/s}$ for $^{12}$CO and $^{13}$CO and $-15$ to $-8$ $\mathrm{km/s}$ for C$^{18}$O.  
Using the optically thin C$^{18}$O, we detected two regions of different velocity peaks (Figure \ref{fig:brcnew}), one coinciding with the IRAS sources and the other 1.3' away from the IRAS sources towards the northwest direction. The first peak can be easily resolved in the velocity range mentioned above. The second peak was resolved in the velocity range $-9$ to $-7$ $\mathrm{km s^{-1}}$.

%??The position-velocity diagrams were plotted along multiple strips of constant galactic longitude. In figure\ref{}?? we show the $C^{18}O$ P-V diagram of constant galactic longitude, which shows two regions with different velocities, suggesting two separate peaks. These are two peaks on the integrated intensity map $C^{18}O$. Figure \ref{??} shows the P-V diagram along the line connecting the ionizing source direction and the cloud; we see a redshift in the velocity as we go away from the bright rim tip towards the tail. This redshift is gradually reduced when we move the strip from one end of the BRC to the southeast direction towards the other and of the BRC.

\section{Discussion}
\label{discuss}
\subsection{Analysis of the properties of the YSOs}
Using optical, NIR, and MIR data, we obtained around 43 YSOs. The age of most of the YSOs is found to be less than 10 million years, and at least one of them is still in its protostellar phase. The masses of the YSOs range from BD to intermediate-mass stars (0.075 M$_\odot$ to 6.4 M$_\odot$). Here, we briefly discuss the results obtained from the SED fitting and spectroscopy of four YSO candidates.

% %\begin{figure}
% {
%     \includegraphics[width = \columnwidth]{brc44_cores.pdf}
%     }

%     \caption{Two clumps detected from C$^{18}$O (integrated intensity) data overplotted on 5.8$\mu$m image. The white contour contains all the YSOs detected in our study using optical, NIR, and MIR wavelengths. The red contour is the new clump detected, in which no YSOs were detected in our study (velocity range : $-9$ to $-7$ $\mathrm{km s^{-1}}$). The contour levels for the red contour are 0.22, 0.37, 0.51, 0.65, and 0.8 K km/ s and those of the white contour are 0.51 K km/s, 1.03, 1.55, 2.08, 2.6 K km/s respectively}
%     \label{fig:brccores}
% \end{figure}

Source S\_0 is seen as an extremely reddened object in both the NIR CC and CM diagrams. This source was studied by \citet{2020ApJ...904...75L} and was categorized as an intermediate-mass star of mass ranging from 0.5 - 12 $\mathrm{M_\odot}$. It does not have an optical counterpart, and the SED fitting shows the mass to lie between 4.4 - 6.4 $\mathrm{M_\odot}$, thus suggesting that it is an intermediate-mass star. 
%We back-traced the source in the NIR CMD, assuming extinction law by \citet{1981ApJ...249..481C}, and obtained the mass of the source around $5_\odot$ and the $A_V$  around 28 mag contrary to that obtained by .... ($A_V$ between 14 and 23 obtained from SED fitting)
Their calculated bolometric luminosity ranges from 75 - $1.1\times 10^4 \mathrm{L_\odot}$. The luminosity that we obtained was $372 - 1172 \mathrm{L_\odot}$. The spectroscopic result shows Bracket gamma emission, which may suggest magnetospheric accretion and the presence of CO bandheads in emission points towards irradiation of the accretion disk by the central source \citep{2010A&A...510A..32M}. All of the results above also suggest that the source is an intermediate-mass star.

S\_1 has both He {\sc ii}, which probes outflowing winds due to accretion, and Paschen emission lines, which probe magnetospheric accretion \citep{Ghosh2023}. This is also an optical source having counterparts in all ten wavelength bands that were used for the SED fitting. The age and mass obtained from the SED fitting are 2 - 3 million years and $\sim$ 3 $\mathrm{M_\odot}$, respectively.

S\_2 is the driving source of the Herbig-Haro object \citep{2002AJ....123.2597O}. The optical CMD predicts its age to be less than a million years, and the SED fitting gives its age to be around a few thousand years, which means that it is a young source. The prediction of the mass of the object by the CMD and SED fitting gives almost the same mass as 0.6 M$_\odot$ (mass range is mentioned in Table.\ref{tab:source_table}). The accretion rate obtained from the SED is of the order of $10^{-6}\mathrm{M_\odot/yr}$, which suggests that it is a highly accreting source. It was identified as a Class~{\sc ii} source using the \citet{2009ApJS..184...18G} criterion. The spectroscopic result shows features in $K$-band. H$_2$ lines, along with CO absorption bands, were detected in this source. %??as opposed to view of (Casali and Matthews 1992) that CO absorption YSOs tend to be class II objects (Casali and Matthews 1992)??.
The presence of H$_2$ emission points to the emission of shock-heated hydrogen molecules caused by outflows or from the photo-dissociation region (PDR; region where photons of energy $<$ 13.6 eV transverse and are not able to ionize but can dissociate molecular species) and CO absorption indicates the presence of a weak circumstellar disk \citep{2010A&A...510A..32M}.
The discrepancy between the youth of this YSO, where SED fitting is pointing towards a very young source, whereas spectroscopic and Spitzer IRAC/ MIPS classification pointing towards a Class {\sc ii} source may be due to the wrong interpretation of the source as a Class {\sc i} source due to excess $24\mu \text{m}$ emission \citep{2009ApJS..184...18G}. 

S\_3 is an optical source with measurements available in all ten bands. %With an age greater than a million years old. 
The SED fitting indicates that it is about 8 million years old. The Paschen$(5-3)$ emission line suggests magnetospheric accretion \citep{Ghosh2023}. The mass obtained by the SED fitting is about 2 $\mathrm{M_\odot}$ and that from optical CMD is $\sim$ 1.9$\mathrm{M_\odot}$; both methods give similar mass estimate. 
Sources S\_1, S\_2, and S\_3 were observed by \citet{2002AJ....123.2597O} as H$\alpha$ emission stars.

\begin{figure}
	\includegraphics[width =\columnwidth]{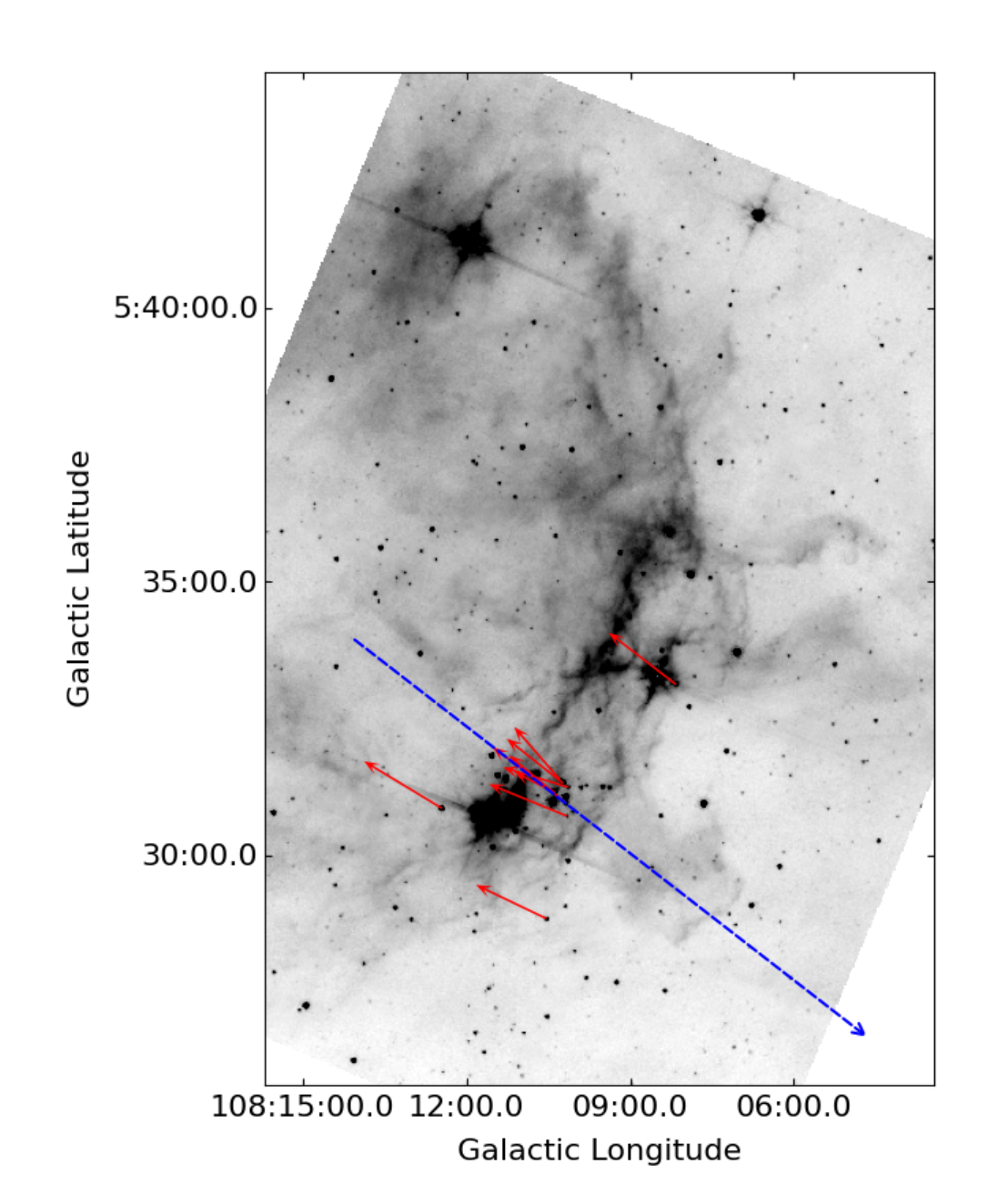}
    \caption{Proper motions of YSO candidates relative to the ionizing source HD 213023 overplotted on Spitzer 5.8 $\mu$m image. The blue dashed line indicates the direction of the ionizing source(s) (towards south-west direction). A clear indication that the YSO candidates of BRC 44 are following an opposite `Rocket motion' with respect to the position of ionizing source(s) can be seen.}
    \label{fig:vectorplot}
\end{figure}
 \begin{figure*}
{
    \includegraphics[width = \textwidth]{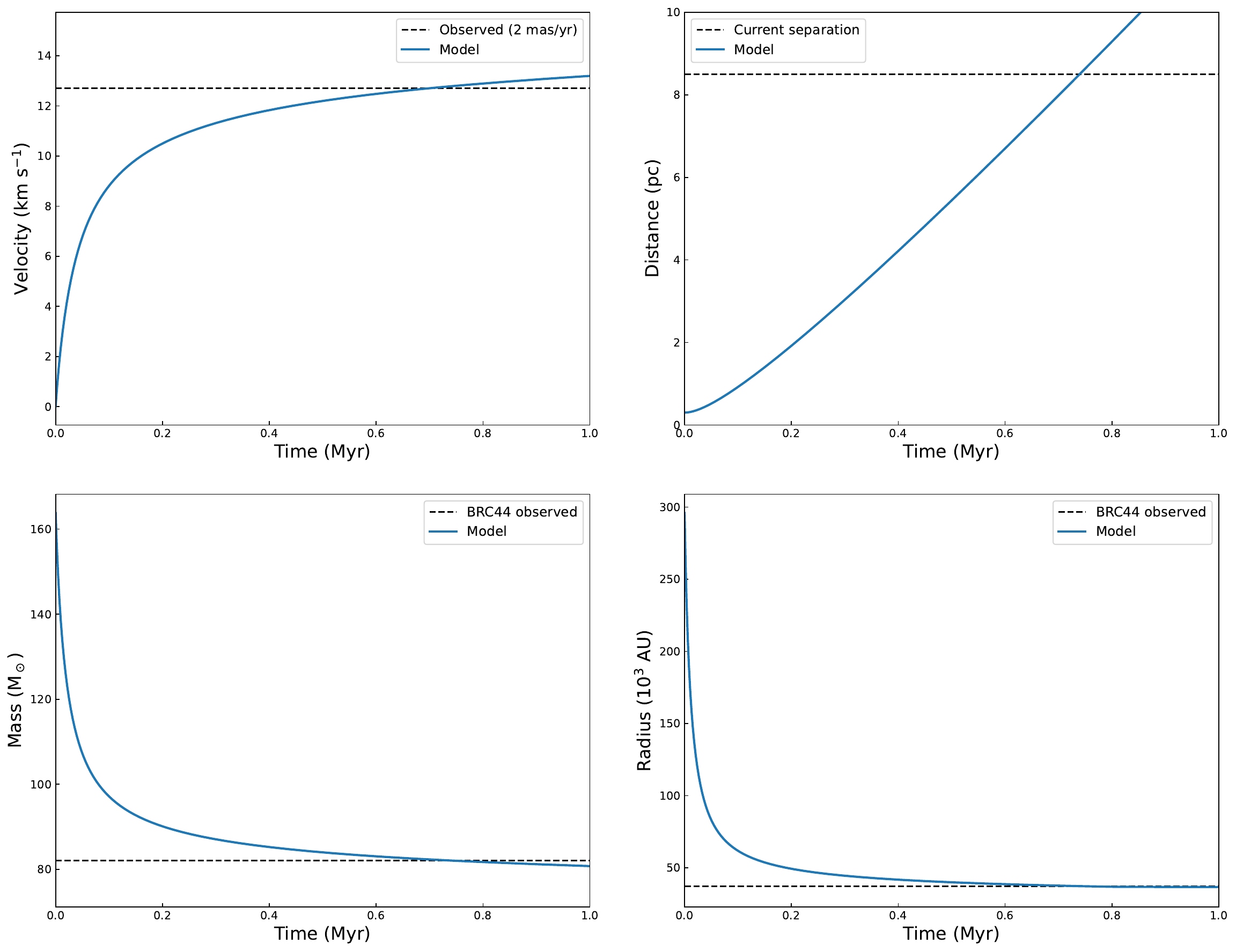}
    }

    \caption{Result of the semi-analytical analysis of RDI and the photoevaporative rocket effect. The image shows velocity, distance, mass, and radius of the cloud from left to right and then top to bottom. The dashed line is the current observed values, and the model recreates it at around 0.7 Myr.}
    \label{fig:rocket_model}
\end{figure*}

\subsection{Cloud Morphology}
When a molecular cloud is influenced by external radiation, the structure and dynamics of the cloud change with respect to the position of the ionizing source, and the radiation ionizes the cloud. Radiation with energy less than 13.6 eV excites or partially ionizes other molecules within the cloud, thus forming different layers beginning with ionized gas, followed by an IBL, PDR, and finally the neutral molecular cloud \citep{2023A&A...677A.152H}. The detailed explanation was given in Section. \ref{struct and dynamics}. 

%20 cm NVSS  traces the free-free emission by the ionized medium at the edge of the molecular cloud in the axis connecting the cloud and ionizing source. In Figure \ref{fig:brcnew}, we can clearly distinguish the IBL traced by 20cm NVSS , PDR, which is traced by Spitzer $8\mu$m emission, and the neutral molecular cloud traced by optically thick $^{12}\mathrm{CO}$. 

\citet{1991ApJS...77...59S} classified BRCs according to the shape of their head as type A, B, and C, representing different stages of evolution. \citet{1994A&A...289..559L} through their simulation showed the evolution of BRC from A to C, where A has a moderate curvature with an `earlike' feature towards the side opposite to that of the illuminated side, B has a little more curved `Elephant Trunk like' structure and C has sharply curved `Cometary' Globule structure. Although not every BRC follows this evolution from type A to type C, every one of them passes through stage A type, and the presence of ear-like feature suggests the early stage where the shock has not yet completely led to cloud collapse \citep{2009ApJ...692..382M}.  BRC 44 shows a type A shape as it has a moderate curvature and an `earlike' feature, as seen in Figure \ref{fig:12co_moment0}. 
This suggests that BRC44 is in its early phase of evolution.

If the density of the cloud is high enough, a D-type shock \citep{2012PhDT.........4T} will propagate through the cloud, perturbing the density of the region. In our region, we compared the external pressure of the ionized medium with the internal pressure of the turbulent gas. 
%We find that internal pressure dominates slightly over the external pressure using $^{13}CO$ line and is less than external pressure using $C^{18}O$ molecular line. Similar analysis by \citet{2006A&A...457..865B} using $^{13}CO$ and $C^{18}O$ revealed that internal pressurThe past, present and future of observations of externally irradiated diskse is less than external pressure thus supporting our result obtained using $C^{18}O$.
We find that for both molecular species, the BRC is under pressure compared to external pressure. This was also concluded by \citet{2006A&A...457..865B} through CO analysis.
For shock propagation to occur, the external pressure due to the IBL should be greater than the internal pressure as analyzed by \citet{2004A&A...426..535M} for many other regions of the BRC. If it were the other way around, the shock could propagate after further ionizing the cloud as more ionized medium will be available for the push. We have only considered the small region that encloses the highly embedded stars for the pressure analysis.

\subsection{Rocket effect in the molecular cloud}
According to the RDI model, the evolution of a remnant cloud containing pre-existing clumps that are photoionized proceeds in two phases: the collapse and the transient phase \citep{1989ApJ...346..735B,1994A&A...289..559L} and the cometary phase \citep{1990ApJ...354..529B,1989ApJ...346..735B}. The collapse phase starts when an
ionization front, preceded by a shock wave, propagates through the
cloud, causing it to compress on its axis, and lasts for $\sim$ 10$^5$ yr. During this phase, a high-density core may form, which could eventually be the potential site for triggered star
formation. At the ionization front, the ionized gas escapes toward the ionizing source, causing a photo-evaporation
flow. Consequently, to conserve momentum, the cloud moves away from the ionizing source \citep{1954BAN....12..177O,2022MNRAS.510.2644S}, which is known as the ``Rocket Effect''.  
In the cometary phase, the cloud undergoes a quasi-equilibrium state, continues accelerating
away from the ionizing source, and may eventually get photo-evaporated completely. %Typically, a B2 star can push a cloud with a speed of $\sim$ 19 km s$^{-1}$ , which eventually goes down to a final speed of 3 - 4 km s$^{-1}$ for a typical cloud having mass of 100 M$\odot$ (Kahn 1954).
The time at which star formation starts and the fraction of
cloud material converted into stars critically depends on the
amount of incident ionizing flux \citep{2011ApJ...736..142B}. Therefore, if star formation is triggered in BRCs, the YSOs that are formed inside the accelerating cloud may share a
similar motion and hence would also move approximately radially
away from the ionizing star(s), which are responsible for the triggering \citep{1990ApJ...354..529B,2022MNRAS.510.2644S}.

To examine the rocket effect in our target region, we used proper motions of the stars from GAIA DR3 data \citep{2022MNRAS.510.2644S}. The vector diagram showing the relative proper motion of the YSOs with respect to the probable ionizing source is shown in Figure \ref{fig:vectorplot}. 
We can see that the projected motion of the YSOs, which implies that the motion of the cloud (assuming the YSOs follow the gravitational potential of the cloud \citep{2011EAS....51...45E}), is clearly away from the axis connecting the ionizing source and BRC. We have to note that in our proper motion analysis we are considering only projected motion of the YSOs and not their 3-dimensional trajectory. 

The proper motion value of the maser source (source A) in Group 2 is available in \citet{2019ApJ...885..131R}, which also has a similar proper motion value and direction similar to the sources in Figure \ref{fig:vectorplot}. So, source A, part of Group 2 sources, also follows the rocket motion.

Now we studied the possible induced motion of the cloud due to the rocket effect using the simple semi-analytical model introduced in \citet{2020MNRAS.497.3351R}. This solves the equation of motion of a spherical globule. BRC 44 is situated near two massive stars, HD 213023 and BD 62 2078 \citep{1994A&A...283..963C}. Among these,  HD~213023 is an O9V star about 8.5 pc away from the bright rim, and another BD 62 2078, an O7.5V star, is about 13.8 pc away from the bright rim \citep{2000AJ....119..323S}. For the analysis we assume clump has mass $\mathrm{M_c}$, radius $\mathrm{R_c}$. The speed of sound in the ionized medium is $\mathrm{c_{ii}}$. The ionizing source is at a distance $D$ and emits $N_{ly}=10^{49}$\,ionizing photons\,s$^{-1}$ (here we consider consider an approximate value of ionizing photons from an O9V spectral type star) \citep{1973AJ.....78..929P}. The equation of motion is given as \citep{1998A&A...331..335M},
\begin{equation}
M_c\ddot D + \dot{M}_c(\dot{D} - \mathrm{C_{ii}}) = 0 .
\end{equation}
The sign of the photo-evaporative flow is taken as negative, as it is in the opposite direction.

The mass loss rate $\dot M_c$, taken from \cite{1998A&A...331..335M} is, 
\begin{equation}
\dot{M}_c = -F' \, \mu \text{m}_H \, \pi \left( \frac{\pi R_c^2}{2} \right) , 
\end{equation}
where,
\begin{equation}
F' = \frac{F_0}{1+\frac{\alpha _ {B} F_{0} R_c}{3\mathrm{C_{ii}}^2}} ,
\end{equation}
is the geometrically diluted ionizing radiation reaching the clump, $\alpha_B$ is the recombination coefficient for hydrogen, and $F_0$ is the unattenuated ionizing flux at the globule considering only geometric dilution from the UV source (i.e. $F_0 = N_{ly}/4/\pi/D^2$).

 The currently observed parameters of the cloud which have to be satisfied by the model are a mass $\sim$ $82\, \mathrm{M_\odot}$ with radius $\sim$ 0.18\,pc, LSR tangential velocity 12.7\,km/s, and distance to the ionizing source $\sim$ 8.5\,pc \citep{2004A&A...426..535M}.

To study the evolution of the cloud size with the mass, we adopted a two-stage process to map the RDI process, followed by a steady evaporation phase of a collapsing cloud \citep{2009MNRAS.398..157H}. During the early phase of compression, that is, the initial 0.8 Myr, the cloud mass reduces by a factor of $\sim$ 2 compared to the factor of $\sim$ 8 reduction in the size of the cloud \citep{2009MNRAS.398..157H}.

 With the above consideration, we found that we can approximately match the current parameters of the cloud for an initial cloud mass $\sim$ 164 $\mathrm{M_\odot}$, radius of $1.44$ pc, zero velocity and distance of $\sim$ 0.3 pc. The evolution of this model is shown in Figure 
\ref{fig:rocket_model}. The time required for the BRC parameter to reach current values is approximately $0.7$ Myr, which means that the rocket effect began around $0.7$ Myr and star formation also began approximately around that time \citep{2011ApJ...736..142B}. We have to keep in mind that these simple models are not expected to exactly fit all parameters of the globule, but give us a rough estimate considering the two-phase evolution of the cloud.
 
 BRC 44 was categorized as having a rim type `A' with an `earlike' feature which points the age of RDI to be very small (less than 1Myr for every type of cloud modeled in \citet{2009ApJ...692..382M}) or in other words the compression phase of the cloud is still going on \citep{2009ApJ...692..382M}. \citet{2006A&A...457..865B} found four sources OVRO 1, OVRO 2, OVRO 3, and OVRO 4, in a 12$''$ vicinity of source A. The
strongest mm source, OVRO 2, is most likely the YSO associated with source A. The dust emission morphology and properties of OVRO 2 suggest that
this IM protostar \citep{2020ApJ...904...75L} is probably in transition between Class 0 and I. So the source A is in its extremely young phase of evolution (also from the analysis done by \citet{1989ApJ...342L..87S} where they compared the dynamical timescale of outflow with the timescale of formation of the shock front). Hence, from the model, the morphology of the cloud, the age of the source A outflow and the pressure imbalance, we can conclude that RDI is taking place in the cloud and it began around a million years ago. The shock from RDI mechanism also may be responsible for the formation of the C$^{18}$O clump that we detected.

Source B/ S\_0, which is considered a part of Group 2, is located at a distance of $\sim 35''$ east of source A, and is likely to be part of a separate group. Its age determined from SED is 5.6 million years.  
Further, the age determined for Group 1 sources (which ranges from 1.2 - 8.2 million years old, of which two sources have an age greater than 5 million years) is also comparable to the age of O-type stars.
This further provides us with evidence that Group 1 YSOs may have been formed at the same time as the O-type star, and that source A in Group 2 may have been triggered. The Group 1 YSOs showing the Rocket effect even though they are not triggered, suggests that they are just following the gravitational potential of the whole cloud, and the rocket acceleration could not overcome this gravitational acceleration \citep{2011EAS....51...45E}.

\section{Conclusions}

In the present work, we studied the BRC~44 region using Spitzer-IRAC/ MIPS observations along with the optical
($V$, $I$) and NIR ($J$, $H$, $K$) data. We identified 43 YSOs present in the star-forming region BRC 44.
Our deep NIR data shows that these young sources have masses ranging from the brown dwarf limit to
intermediate-mass stars.  The
approximate ages of YSOs were determined from SED fitting and optical/NIR CMD. We noticed two groups
of YSOs of different ages in the region. The age of Group 1 YSOs
turns out to be a few million years, comparable to the age of nearby ionizing star(s). The NIR spectroscopy observations of a couple of YSOs confirm that these are in the early evolutionary stages.
We further studied the cloud using millimeter CO data. The cloud mass and pressure were measured, and was found that the cloud
is under-pressured compared to IBL pressure, suggesting that RDI may be the possible
mechanism for star formation in BRC 44. The cloud morphology was studied from the $^{12}$CO data to get an idea of the youth of the cloud, and it was found from both $^{12}$CO and using a semi-analytical approach of RDI, the BRC was formed approximately 0.7 Myr ago. The projected motions of the YSOs with respect to the ionizing star(s) suggest ongoing rocket motion in the region. 

We infer that Group 1 YSOs may be formed at the same time as the ionizing source and were pushed along with the cloud as the ionizing rays started to photoionize the cloud and the RDI scenario came into effect. And after a time period, triggered star formation was initiated by the shock front, which resulted in the formation of source A. So the whole BRC 44 region was already a site for stars to form, the RDI mechanism just triggered new episodes of star formation, and the cloud structure and dynamics were further altered due to this feedback.

\section*{Acknowledgements}
We thank the anonymous referee for the constructive comments and suggestions that significantly improved the manuscript. 
NP and RC acknowledge the financial support received through the SERB CRG/2021/005876 grant. 
TJH acknowledges UKRI guaranteed funding for a Horizon Europe ERC consolidator grant (EP/Y024710/1) and a Royal Society Dorothy Hodgkin Fellowship. D.K.O. acknowledges the support of Department of Atomic Energy, Government of India, under Project Identification No. RTI 4002. RKY gratefully acknowledges the support from the Fundamental Fund of Thailand Science Research and Innovation (TSRI) (Confirmation No. FFB680072/0269) through the National Astronomical Research Institute of Thailand (Public Organization).
This publication makes use of data from the Two Micron All Sky Survey (a joint project of the University of Massachusetts and the Infrared Processing 
and Analysis Center/ California Institute of Technology, funded by the 
National Aeronautics and Space Administration and the National Science 
Foundation), archival data obtained with the {\it Spitzer Space Telescope}
(operated by the Jet Propulsion Laboratory, California Institute of Technology, under contract with NASA). This work has made use of data from the European Space Agency mission {\it Gaia} (https://www.cosmos.esa.int/gaia), processed by the {\it Gaia} Data Processing and Analysis Consortium (DPAC). Funding for the DPAC
has been provided by national institutions, in particular, the institutions participating in the {\it Gaia} Multilateral Agreement.

This research made use of the data from the Milky Way Imaging Scroll Painting (MWISP) project, which is a multi-line survey in $^{12}\mathrm{CO}$/$^{12}\mathrm{CO}$/C$^{18}$O along the northern galactic plane with PMO-13.7m telescope. We are grateful to all the members of the MWISP working group, particularly the staff members at PMO-13.7m telescope, for their long-term support. MWISP was sponsored by National Key R\&D Program of
China with grants 2023YFA1608000 \& 2017YFA0402701 and by CAS Key Research Program of Frontier Sciences with grant QYZDJ-SSW-SLH047.

\appendix \label{appendix}
\restartappendixnumbering

\section{Identification of YSOs} \label{identify_spitzer_app}

%Here we discuss in detail how YSOs were identified using  NIR and MIR data.
\subsection{Using NIR data from 2MASS and TIRCAM2}

\begin{figure*}
    \centering
    \subfigure[]{
        \includegraphics[width = 0.48\columnwidth]{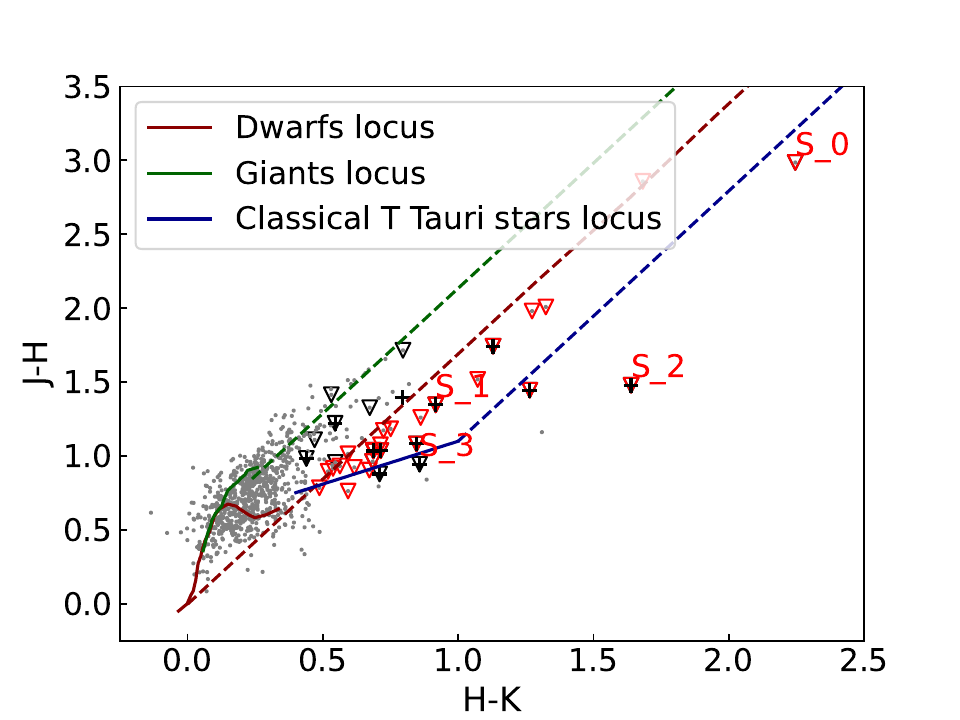}
        \label{a)}
    }
    \subfigure[]{
        \includegraphics[width = 0.48\columnwidth]{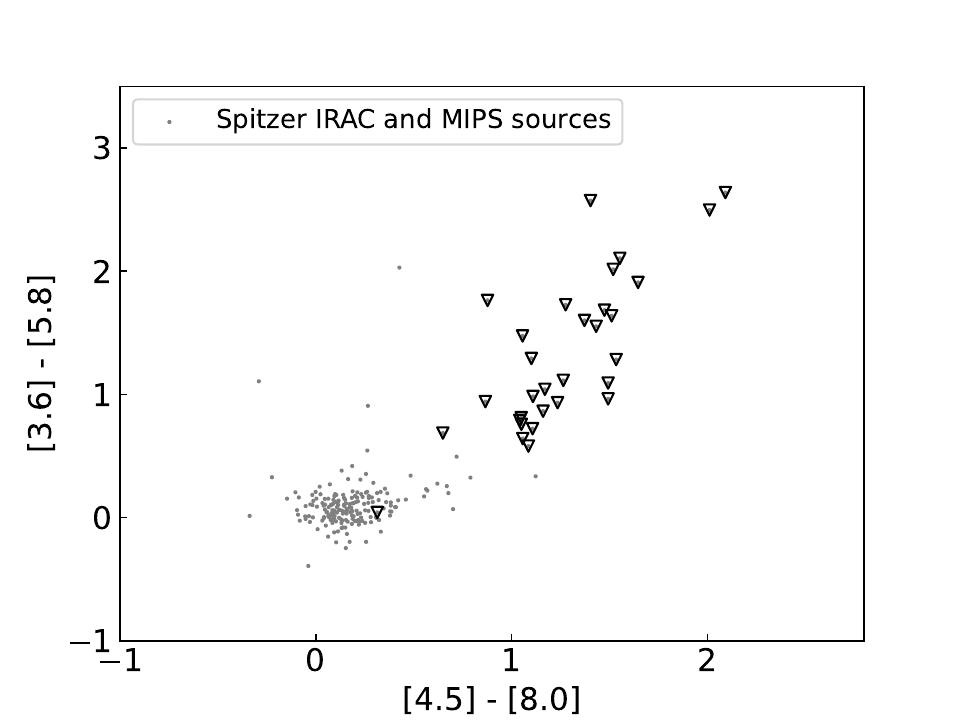}
        \label{b)}
    }
    \caption{a) ($J$ - $H$) vs ($H$ - $K$) color-color diagram for all the sources (grey dots) in the BRC 44 region. We have also overplotted H$\alpha$ stars with `+' marker \citep{2002AJ....123.2597O}, YSOs obtained from NIR excess as red inverted triangle, and YSOs obtained from Spitzer-IRAC and MIPS photometry as black inverted triangle. b) [3.6] - [5.8] vs [4.5] - [8.0] color-color diagram for all the Spitzer-IRAC/ MIPS detected sources (grey dots) in the BRC 44 region. YSOs obtained from Spitzer-IRAC and MIPS photometry are shown as black inverted triangle.}
    \label{fig:cc}
\end{figure*}

The NIR color-color (CC) diagram is an excellent tool for identifying young stellar sources exhibiting the IR excess. In the present work, we used the NIR data obtained using TIRCAM2 and 2MASS. The $JHK$ magnitudes, initially in the 2MASS system, were converted to the California Institute of Technology system using the equations available on the 2MASS website. The ($J$-$H$) vs. ($H$-$K$) CC diagram for all the sources in the region is shown in Figure \ref{fig:cc}(a). The intrinsic locii of dwarfs, giants, and Classical T-Tauri stars (CTTS) are also overplotted. The dashed lines are the reddening vectors drawn from the tip of the M-dwarf, A0V star, and the CTTS locus. We classified sources into three regions in the NIR-CC diagrams \citep[for details; see][]{2017MNRAS.468.2684P}. In brief, sources located between the leftmost and middle reddening vectors are considered to be
either field stars (MS stars, giants) or Class {\sc iii} sources or Class {\sc ii} sources with small NIR
excesses. Sources located between the middle and right reddening vectors are considered
to be mostly CTTS (Class {\sc ii} objects), while sources located redward of the right-most
reddening vector are most likely Class {\sc i} objects (protostar-like objects). Therefore,
the objects lying rightward of the middle reddening vector and above the CTTS locus in the NIR-CC diagrams are considered NIR excess
sources and are probable members of the region. We obtained about 12 sources with NIR excess that were not previously identified as young sources. Since young stars are also H$\alpha$ emitters, we have
identified the counterparts of the H$\alpha$ emission stars from \citet{2002AJ....123.2597O} in our NIR data. The H$\alpha$ emission stars are shown with `+' symbols in Figure \ref{fig:cc}(a).

\subsection{Using Spitzer-IRAC and MIPS data}
Young stellar sources can be identified and classified based on their NIR and Mid-Infrared (MIR) colors from the Spitzer-IRAC and MIPS observations. Other contaminants may mimic the colors of the YSOs in the IRAC CC diagrams. These contaminants include star-forming galaxies, broad-line AGNs, unresolved knots of shock emissions, and resolved structured PAH emission, which contaminates the photometric apertures of some dim field stars, leading to spurious excess emission in the 5.8 and 8.0 $\micron$ bandpasses. These sources were rejected using the criteria discussed in \citet{2009ApJS..184...18G}.

We followed the constraints outlined by \citet{2009ApJS..184...18G} to identify young stellar candidates. After filtering out contaminants from our data, we classified those sources as Class \textsc{I} YSOs that meet the following criteria:
 $[4.5] - [5.8] > 0.7$ and $[3.6] - [4.5] > 0.7$; 

From the remaining Class \textsc{I} sources are differentiated using the following criteria:

$[4.5] - [8.0] - \sigma_1 > 0.5$

$[3.6] - [5.8] - \sigma_2 > 0.35$

$[3.6] - [5.8] + \sigma_2 <= (0.14/0.04)\times (([4.5] - [8.0] - \sigma_1) - 0.5) + 0.5$

$[3.6] - [4.5] - \sigma_3 > 0.15$

Where $\sigma_1$ is an error in ([4.5] - [8.0]) color, $\sigma_2$ is an error in ([3.6] - [5.8]), and $\sigma_3$ is an error in ([3.6] - [4.5]) that are calculated using the errors in corresponding magnitudes.

Using 24 $\micron$ photometry along with IRAC photometry, we identified transition disk sources in the region, which are also marked in the CC diagram in Figure \ref{fig:cc}b. 

 Finally, using the magnitudes available in 3.6 $\mu \text{m}$, 4.5 $\mu \text{m}$, 5.8 $\mu \text{m}$, 8 $\mu \text{m}$ and 24 $\mu \text{m}$, we identified around 37 YSOs using the \citet{2009ApJS..184...18G} criterion. In total, we identified around 49 YSO candidates in the region. The spatial distribution of the identified YSOs is shown in Figure \ref{fig:brcnew}. We notice that the YSOs are mostly concentrated into two groups, one near the bright rim (sources with optical counterparts; labeled as Group 1) and the other farther from the rim (consisting of deeply embedded sources, Group 2). 

 \section{Proper motion and distance estimation}
 \label{distance_app}
\begin{figure*}
    \centering
    \subfigure[]{
        \includegraphics[width = 0.48\columnwidth]{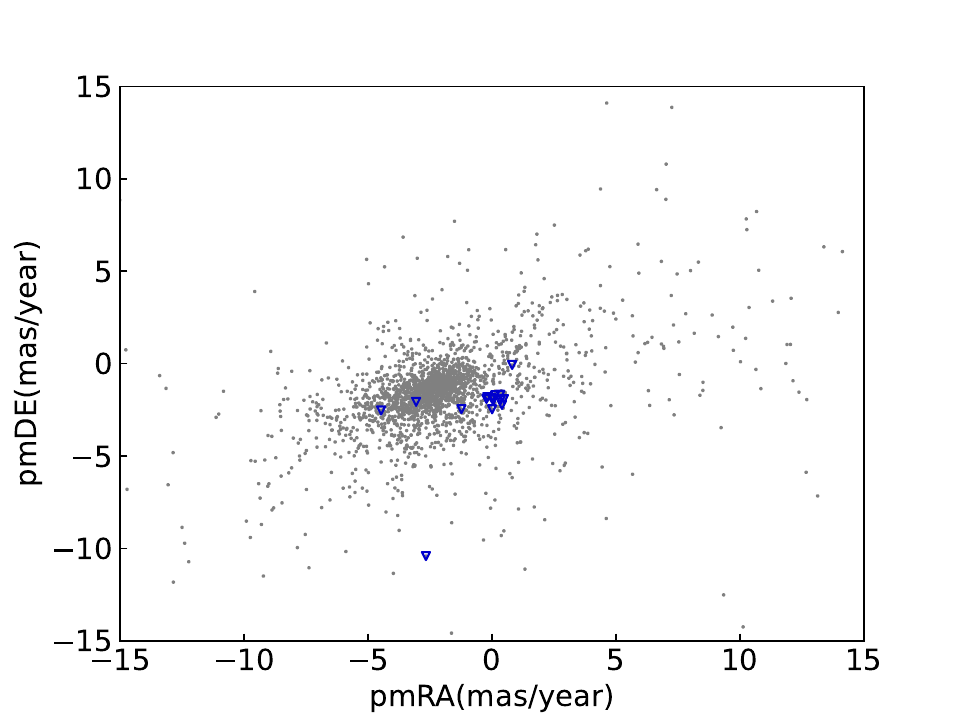}
        \label{fig:propermotion}
    }
    \subfigure[]{
        \includegraphics[width = 0.48\columnwidth]{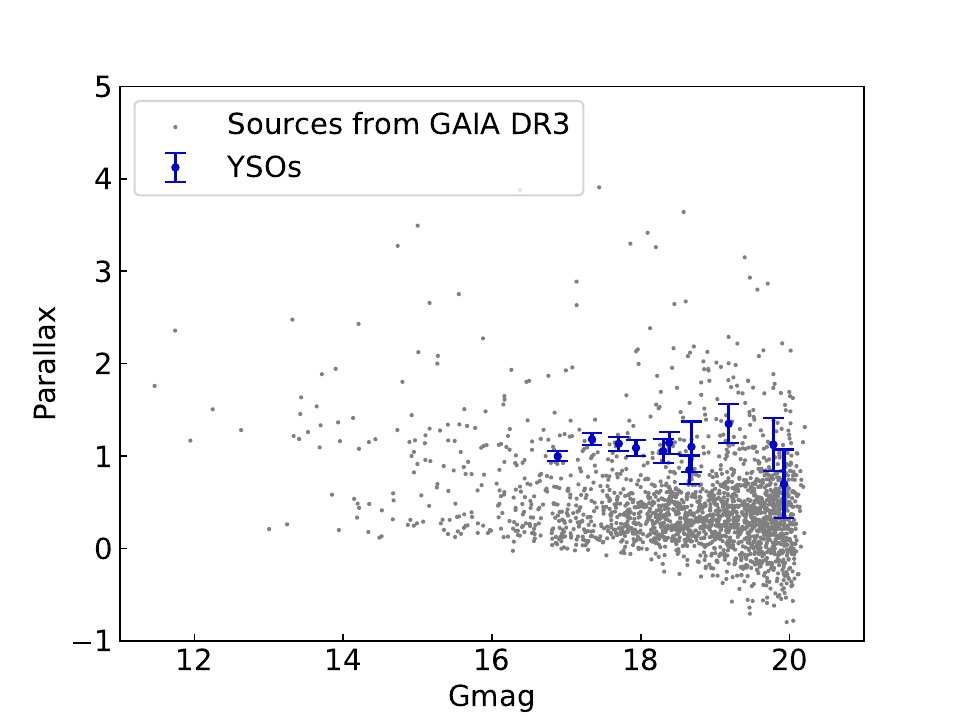}
        \label{fig:parallax}
    }
    \caption{a) Vector point diagram for the sources in the BRC~44 region. The proper motions of the YSO candidates are shown with blue inverted triangles.
    b) Parallax vs $G$-band magnitude for stars in BRC~44 using Gaia DR3. Blue circles show YSOs selected via proper motions.}
\end{figure*}

 The vector point diagram (VPD), constructed by plotting the proper motions in right ascension ($\mu_\alpha$) and declination ($\mu_\delta$), is shown in Figure \ref{fig:propermotion}. It displays the distribution for all stars in the region (gray dots) and the identified YSOs (blue inverted triangles). An overdensity is clearly visible, with the majority of YSOs clustering around ($\mu_\alpha$, $\mu_\delta$) $\sim$ (0.13, –2) mas/yr. To ensure that the selected YSOs are genuinely associated with the region, we applied a 2.5$\sigma$ selection criterion based on the average proper motion and parallax, focusing on those located near the bright rim. This analysis excluded 6 out of 49 YSOs as outliers, leaving 43 candidate YSOs, including 22 newly identified sources not reported by \citet{2009ApJS..184...18G} (some are shown in Table. \ref{tab:photometry}). After excluding the outliers, the mean proper motion of the remaining YSOs was determined to be $\mu_\alpha$ $\sim$ 0.13 $\pm$ 0.07 mas/yr and $\mu_\delta$ $\sim$ –1.95 $\pm$ 0.06 mas/yr.
%???We then plotted a proper motion graph where we plotted pmra$\times$cos(dec) (proper motion along RA) vs pmdec (proper motion along DEC). As can be seen in the diagram, the positions of the YSOs have an overdensity at a range of proper motions. \citep{??}
%After removing these outliers, the mean proper motion of the YSOs is calculated as $\mu_\alpha$ $\sim$  $0.13 \pm 0.07$ mas/yr and $\mu_\delta$ $-1.95 \pm 0.06$ mas/yr. %Finally, the mean proper motion of the YSOs was calculated to be equal to $1.95 \pm 0.07 mas/yr$. 

% \begin{figure*}
% \centering
% \includegraphics[width=0.49\linewidth]{propermotion.pdf}
% \includegraphics[width=0.49\linewidth]{parallax.pdf}
% \caption{Left: Vector point diagram for the sources in the BRC~44 region. The proper motions of the YSO candidates are shown with blue inverted triangles. Right: Parallax vs $G$-band magnitude for stars in BRC~44 using Gaia DR3. Blue circles show YSOs selected via proper motions.}
% \label{fig:parallax}
% \end{figure*}

Several distance estimates were made for this region. \citet{1974PDAO...14..283C} estimated a distance of 0.9 kpc using the absolute magnitude of the nearby O-type star. However,  using the $^{13}\mathrm{CO}$ observations, \citet{1986A&A...169..281C} estimated a distance of $\sim$ 0.8 kpc for Source A. Using the trigonometric parallax of the 6.7 GHz methanol maser, \citet{2010A&A...511A...2R} found the same distance to Source A. In the present work, we used the parallax values of the YSO candidates from the Gaia DR3 catalog and estimated the distance to the region. We plotted the distribution of the parallaxes as a function of $G$-magnitudes, as shown in Figure \ref{fig:parallax}. It clearly shows two distributions of the parallax values. One is around the parallax value $\sim$ 0.5 mas, and the other is around parallax $\sim$ 1 mas. Most of the YSO candidates that were selected as members based on the proper motions have parallax values around $\sim$ 1 mas (after correcting the parallax zero point value of 0.017 mas). We obtained the mean parallax values of the YSO candidates and estimated the distance of the region. We obtained a distance of $\sim 952 \pm 52$ pc to the region, which is comparable to the value of $900$ pc obtained by \citet{1995yCat..41130325H}. This also suggests that most of the YSO candidates are highly likely to be members of the BRC.

\begin{deluxetable*}{llccccccccccc}
\tablecaption{\textbf{Photometric measurements of YSO candidates identified in the BRC region. The entire table is available in machine-readable form}.}
\label{tab:photometry}
\tablehead{
\colhead{ID} & \colhead{$\alpha_{2000}$} & \colhead{$\delta_{2000}$} & \colhead{$V$} & \colhead{$I$} & \colhead{$J$} & \colhead{$H$} & \colhead{$K_S$} & \colhead{[3.6]} & \colhead{[4.5]} & \colhead{[5.8]} & \colhead{[8.0]} & \colhead{[24]} \\
 &  &   &$\pm$ $eV$ &  $\pm$ $eI$ & $\pm$ $eJ$ & $\pm$ $eH$ & $\pm$ $eK$ & $\pm$ $e[3.6]$ & $\pm$ $e[4.5]$& $\pm$ $e[5.8]$ & $\pm$ $e[8.0]$ & $\pm$ $e[24]$ \\
}
\startdata
1 & 22:29:03.48 & +64:14:11.31 & 18.108 & 14.760 & 12.317 & 10.827 & 10.262 & 9.896 & 9.921 & 9.854 & 9.607 & 6.952 \\
    &   &   & $\pm$0.031 & $\pm$0.031 & $\pm$0.036 & $\pm$0.031 & $\pm$0.023 & $\pm$0.078 & $\pm$0.075 & $\pm$0.076 & $\pm$0.060 & $\pm$0.123 \\
2 & 22:28:48.70 & +64:13:28.60 & --- & --- &---& 22.105 & 15.755 & 13.884 & 13.006 & 12.121 & 12.129 & --- \\
    &   &   &   &   &   & $\pm$0.011 & $\pm$0.009 & $\pm$0.089 & $\pm$0.070 & $\pm$0.066 & $\pm$0.096 &\\
3 & 22:28:43.53 & +64:13:18.92 & 19.210 & 16.034 & 14.519 & 13.437 & 12.716 & 12.095 & 11.467 & 11.408 & 10.818 & --- \\
    &   &   & $\pm$0.045 & $\pm$0.035 & $\pm$0.010 & $\pm$0.003 & $\pm$0.009 & $\pm$0.101 & $\pm$0.084 & $\pm$0.075 & $\pm$0.069 &   \\
%4 & 22:28:45.48 & +64:13:22.80 & --- & --- & 15.388 & 14.480 & 13.737 & 12.834 & 12.376 & 11.850 & 11.267 & --- \\
%    &   &   &   &   & $\pm$0.008 & $\pm$0.006 & $\pm$0.010 & $\pm$0.118 & $\pm$0.095 & $\pm$0.103 & $\pm$0.189 &   \\
\enddata
\end{deluxetable*}

\newpage

\bibliography{cite}{}
%\bibliographystyle{aasjournalv7}



\end{document}